\documentclass[journal,twoside]{IEEEtran}
\usepackage{amsmath, graphicx, multirow}
\usepackage{subfigure}
\usepackage{threeparttable,cite}
\usepackage{array}
\usepackage[colorlinks,linkcolor=red]{hyperref}
\usepackage{color}
\begin{document}
	
	\title{Context Adaptive Extended Chain Coding for Semantic Map Compression}
	\author{
		Runyu Yang, \IEEEmembership{Member, IEEE},
		Junqi Liao, \IEEEmembership{Graduate Student Member, IEEE},
		Hyomin Choi, \IEEEmembership{Member, IEEE},
		Fabien Racap\'e, \IEEEmembership{Member, IEEE},
		and~Ivan V. Baji\'c, \IEEEmembership{Senior Member, IEEE}
		\thanks{
			Date of current version \today. 
			
			Runyu Yang and Ivan V. Baji\'c are with the School of Engineering Science, Simon Fraser University, Burnaby, BC V5A 1S6, Canada (e-mail: runyuy@sfu.ca; ibajic@ensc.sfu.ca).
			
			Junqi Liao is with the CAS Key Laboratory of Technology in Geo-spatial Information Processing and Application System, Department of Electronic Engineering and Information Science, University of Science and Technology of China, Hefei 230027, China (e-mail: liaojq@mail.ustc.edu.cn).
			
			Hyomin Choi and Fabien Racap\'e are with AI Lab, InterDigital, Los Altos, CA (email: hyomin.choi@interdigital.com; fabien.racape@interdigital.com).
		}
	}
	
	\markboth{Submitted to IEEE Transactions on Image Processing}
	{Yang \MakeLowercase{\textit{et al.}}: Context Adaptive Extended Chain Coding for Semantic Map Compression}
	\maketitle
	
	\begin{abstract}
		Semantic maps are increasingly utilized in areas such as robotics, autonomous systems, and extended reality, motivating the investigation of efficient compression methods that preserve structured semantic information. This paper studies lossless compression of semantic maps through a novel chain-coding-based framework that explicitly exploits contour topology and shared boundaries between adjacent semantic regions. We propose an extended chain code (ECC) to represent long-range contour transitions more compactly, while retaining a legacy three-orthogonal chain code (3OT) as a fallback mode for further efficiency. To efficiently encode sequences of ECC symbols, a context-adaptive entropy coding method based on Markov modeling is employed. Furthermore, a skip-coding mechanism is introduced to eliminate redundant representations of shared contours between adjacent semantic regions, supporting both complete and partial skips via run-length signaling. Experimental results demonstrate that the proposed framework achieves an average bitrate reduction of 20\% compared with a state-of-the-art benchmark on semantic map datasets. Extended evaluations on occupancy maps further confirm compression gains across the majority of tested scenarios. The source code is made publicly available at \url{https://github.com/InterDigitalInc/LosslessSegmentationMapCompression}.
	\end{abstract}

	
	\section{Introduction}
	\label{sec:intro}
	
	\begin{figure}
		\centering
		\begin{minipage}[b]{0.7\linewidth}
			\centering
			\centerline{\includegraphics[width=\linewidth]{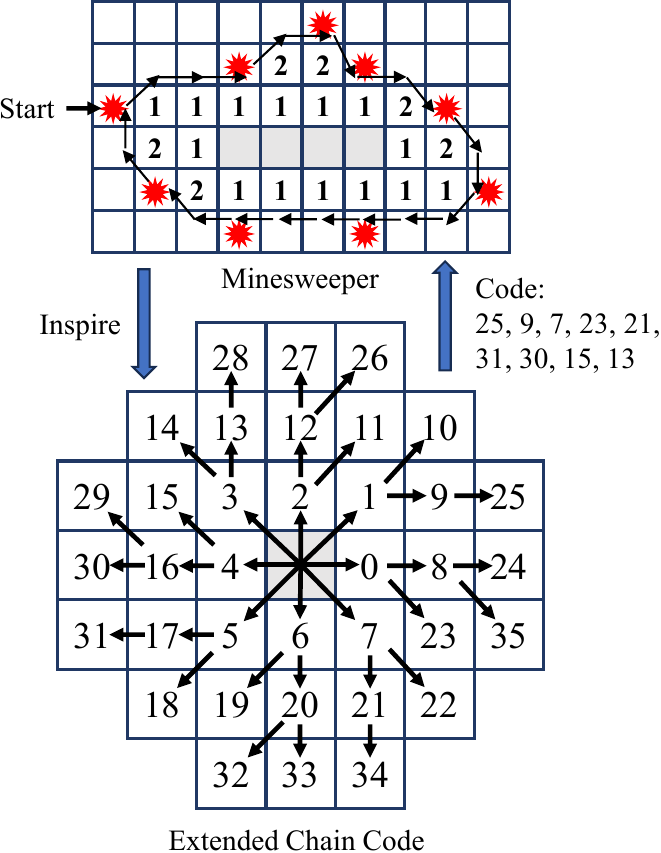}}
		\end{minipage}
		\caption{Symbols of extended chain code and its usage for representing the Minesweeper contour. In the top map, the mine is indicated by the asterisk, and the value shows the number of mines in 8-neighbourhoods.}	
		\label{fig:fig1}
	\end{figure}
	
	Semantic maps provide structured interpretations of visual content and are widely studied and utilized in robotics, autonomous systems, and extended reality applications. In extended reality systems, semantic understanding of the environment can be computed on a device or edge server and conveyed to user devices to support human--environment interaction, scene understanding, or spatial anchoring. Similarly, in autonomous driving and robotic perception, semantic maps are used to organize scene elements and may contribute to the construction of structured 3D representations of the environment. In these scenarios, semantic maps often exhibit strong geometric and topological structure, motivating the study of compression methods that can exploit such properties.
	
	In many image and video coding systems, semantic maps are already compressed for storage, transmission, or use as intermediate representations. In some applications, exact reconstruction of semantic maps is desirable. For example, semantic maps have been used as intermediate representation layers to enable high-fidelity image reconstruction in image compression frameworks \cite{akbari2019dsslic}. In such systems, generic lossless image codecs, such as FLIF \cite{sneyers2016flif}, are commonly employed. Li \emph{et al.} \cite{li2026moric} further leveraged semantic maps to achieve region-based implicit codec for image compression, compressing the maps using chain coding based contour coding. Semantic maps also play an essential role in point cloud compression. In video-based point cloud compression (V-PCC), the state-of-the-art dynamic point cloud coding standard developed by MPEG \cite{preda2017report}, a semantic map known as the occupancy map indicates whether a projected 2D pixel corresponds to a valid 3D point. Within the V-PCC framework, the screen content coding (SCC) extension of the High Efficiency Video Coding (HEVC) standard \cite{xu2015overview} is used to compress occupancy maps. While these approaches benefit from mature general-purpose compression tools, they do not explicitly exploit the structural properties of semantic maps, which limit compression efficiency.
	
	\begin{figure*}[th!]
		\centering
		\begin{minipage}[b]{0.9\linewidth}
			\centering
			\centerline{\includegraphics[width=\linewidth]{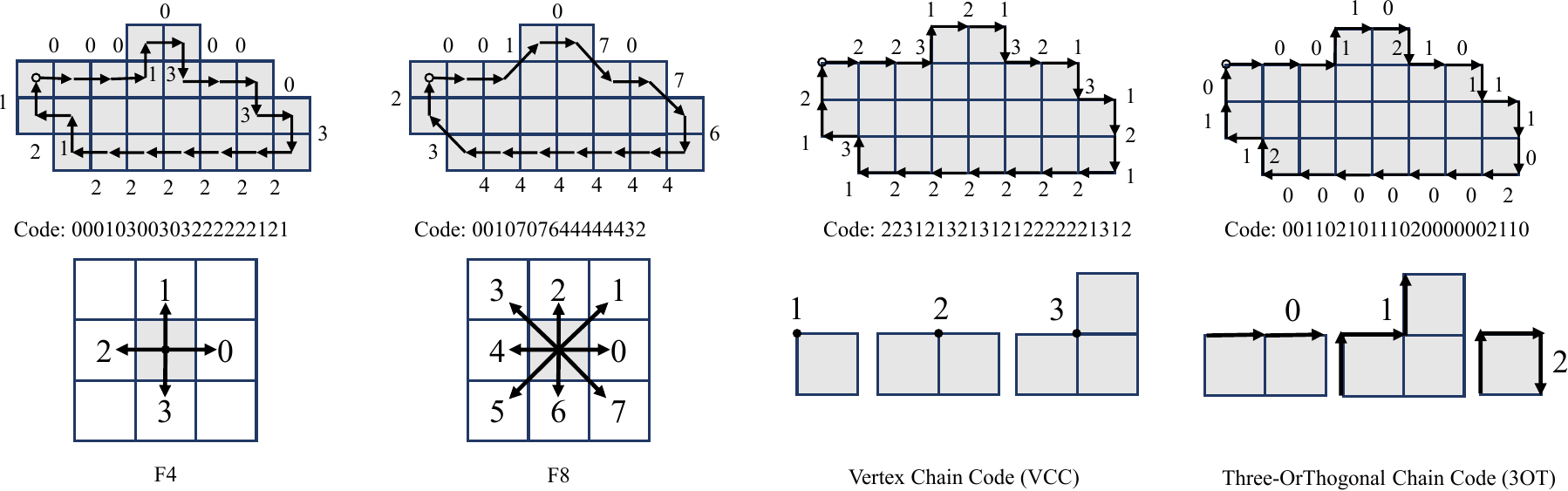}}
		\end{minipage}
		\caption{Examples of classical chain code representations for contours: Freeman F4 and F8 chain codes, VCC, and 3OT.}	
		\label{fig:F84VCC3OT}
	\end{figure*}
	
	A semantic map typically exhibits two distinctive characteristics: a limited number of discrete values and sharp boundaries between neighboring regions. Despite these well-defined structural properties, relatively few compression techniques have been specifically designed and used for semantic maps. Most existing approaches rely on universal image compression methods, such as those discussed earlier, which are not optimized to handle the combination of limited value cardinality and sharp inter-region boundaries, often resulting in suboptimal compression efficiency.
	
	Chain coding is a classical and effective technique for representing object contours and is therefore well suited to semantic map compression. Since its introduction by Freeman in 1961 \cite{freeman1961encoding}, chain coding has undergone extensive development, including the Freeman four-direction (F4) \cite{freeman1974computer} and eight-direction (F8) chain codes, as well as more compact representations such as the vertex chain code (VCC) \cite{bribiesca1999new} and the three-orthogonal chain code (3OT) \cite{cruz2005compressing}. Numerous entropy coding techniques have been applied to compress chain code sequences, including chain differences \cite{freeman1974computer}, Huffman coding \cite{liu2005efficient}, Markov models \cite{lu1991highly}, and adaptive arithmetic coding \cite{chan1995highly}. However, most existing chain coding methods were originally designed for binary or bi-level images and do not directly address the multi-value nature of modern semantic maps.
	
	To extend chain coding to semantic map compression, Yang \emph{et al.} \cite{yang2020chain,yang2024cc} proposed combining chain coding with quadtree-based block partitioning. While this approach significantly improves compression efficiency compared with generic image codecs, the use of quadtree partitioning introduces two fundamental limitations. First, partitioning truncates chain code sequences, which degrades contextual modeling and reduces the effectiveness of entropy coding. Second, partition signaling introduces non-negligible overhead, which becomes particularly inefficient for full-resolution occupancy maps with many fine structures.
	
	In this paper, we propose a chain-coding-based semantic map compression framework that eliminates the need for block-based partitioning and represents each semantic region exclusively through its contour. Our study introduces a novel extended chain code (ECC) that improves representational efficiency by encoding longer contour transitions using an expanded symbol set. The illustration of ECC is shown in Fig.~\ref{fig:fig1}. It is inspired by the Minesweeper game, which only requires a small number of mines to represent the contour. The framework also retains the three-orthogonal chain code (3OT) as a fallback mode for further efficiency. To efficiently encode sequences of ECC and 3OT symbols, a Markov model-based entropy coding method is employed. Furthermore, we introduce a skip-coding mechanism that explicitly exploits shared boundaries between adjacent semantic regions, using complete skip mode with run-length coding to avoid redundant contour representations when contours overlap with previously encoded regions.
	
	The main contributions of this paper are summarized as follows:
	\begin{itemize}
		\item A chain-coding-based semantic map compression framework that represents semantic regions solely through their contours without block-based partitioning.
		\item An extended chain code (ECC) with an expanded symbol set to improve contour representation efficiency, together with a three-orthogonal chain code (3OT) used as a fallback mode.
		\item A skip-coding mechanism with run-length signaling to eliminate redundant representations of shared contours between neighboring semantic regions.
		\item Experimental results demonstrate approximately 20\% bitrate savings on average over a state-of-the-art compression framework CC-SMC \cite{yang2024cc}.
	\end{itemize}
	
	\begin{figure*}[th!]
		\centering
		\begin{minipage}[b]{1\linewidth}
			\centering
			\centerline{\includegraphics[width=\linewidth]{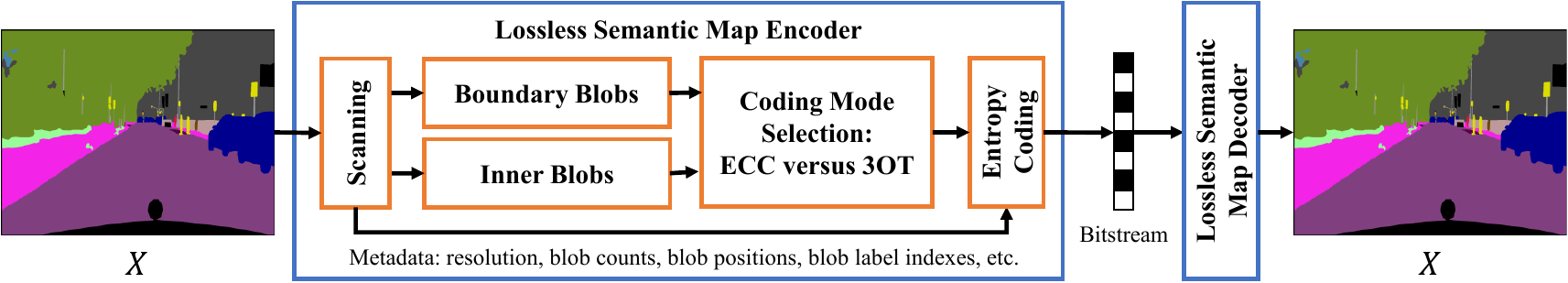}}
		\end{minipage}
		\caption{System schematic of the overall lossless compression framework using the proposed extended chain coding (ECC) method.}	
		\label{fig:overview}
	\end{figure*}
	
	\section{Related Work}
	
	\subsection{Chain Coding Methods}
	
	Chain codes are a classical representation for encoding contours and rasterized object boundaries. They were first introduced by Freeman \cite{freeman1961encoding}, who proposed the four-direction (F4) and eight-direction (F8) chain codes based on 4- and 8-neighborhood connectivity, respectively \cite{freeman1974computer}. In these representations, a contour is encoded as a sequence of directional symbols that describe the relative movement between successive boundary pixels.
	
	Subsequent research focused on improving the compactness and statistical efficiency of chain code representations. The vertex chain code (VCC) \cite{bribiesca1999new} encodes local boundary configurations at vertices, while the three-orthogonal chain code (3OT) \cite{cruz2005compressing} reduces the symbol alphabet to three symbols by representing direction changes relative to the previous contour direction. These representations enable more efficient entropy coding while preserving lossless contour reconstruction. Illustrative examples of F4, F8, VCC, and 3OT are shown in Fig.~\ref{fig:F84VCC3OT}.
	
	To further reduce redundancy in chain code sequences, various statistical coding techniques have been explored. Chain difference coding \cite{freeman1974computer, liu2005efficient} exploits correlations between successive symbols, while Huffman coding \cite{liu2005efficient}, Markov models \cite{lu1991highly}, and adaptive arithmetic coding \cite{chan1995highly} have been employed to model symbol probabilities more accurately. Extensions to primitive chain codes have also been proposed by introducing compound symbols that replace frequently occurring symbol patterns. For example, Liu \emph{et al.} \cite{liu2007compressed} extended VCC by adding symbols to represent common symbol pairs, and Zahir \emph{et al.} \cite{zahir2007new} introduced fixed-length directional patterns for difference-coded F8 chain codes. More recently, \v{Z}alik \emph{et al.} \cite{vzalik2021lossless} proposed a lossless compression pipeline for chain code sequences combining the Burrows--Wheeler transform, move-to-front transform, zero-run coding, and adaptive arithmetic coding.
	
	Although these methods significantly improve the compression efficiency of contour representations, they are primarily designed for individual binary or bi-level contours and do not explicitly address redundancy arising from shared boundaries between multiple adjacent regions in multi-value maps.
	
	\subsection{Semantic Map Compression}
	
	Semantic map compression has received increasing attention in recent years, particularly in the context of image and point cloud coding. Early approaches typically relied on universal image compression methods, which treat semantic maps as ordinary images and do not explicitly exploit their region-based structure.
	
	To incorporate contour-based representations, Yang \emph{et al.} \cite{yang2020chain} proposed combining chain coding with quadtree-based block partitioning for semantic and occupancy map compression. In this framework, chain coding is applied within partitioned blocks that satisfy specific conditions, while other blocks are encoded directly. This approach substantially improves compression efficiency compared with generic image codecs. Building upon this work, CC-SMC \cite{yang2024cc} enhanced context modeling by incorporating information from adjacent blocks and extended the framework to inter-frame coding for dynamic segmentation maps.
	
	Despite these improvements, block-based partitioning introduces inherent limitations. Partition boundaries truncate chain code sequences, which reduces the effectiveness of context-based entropy coding. In addition, partition signaling incurs non-negligible bitrate overhead, particularly for full-resolution maps with fine-grained structures. Alternative approaches that apply chain coding to multi-level images by treating each value independently, such as the method proposed by Jeromel \emph{et al.} \cite{jeromel2020efficient} for cartoon image compression, avoid block partitioning but inevitably encode duplicate contours along shared region boundaries, leading to reduced coding efficiency.
	
	These limitations motivate contour-based semantic map compression frameworks that preserve long-range contour continuity and explicitly exploit shared boundaries between adjacent regions without relying on block-based partitioning.
	
	\section{Proposed Framework}
	\label{sec:method}
	
	To preserve long-range contour continuity and to explicitly exploit shared boundaries between adjacent regions, contour representations beyond purely local, unit-step descriptions can be considered when designing semantic map compression frameworks. Conventional chain codes encode contours as sequences of elementary directional moves, which can result in long symbol sequences for extended boundaries. Motivated by this observation, the proposed framework introduces an extended chain code (ECC) that explicitly encodes longer contour transitions.
	
	An overview of the proposed semantic map compression framework is illustrated in Fig.~\ref{fig:overview}. The proposed framework performs lossless compression, ensuring that the reconstructed semantic map is identical to the original input. Section~\ref{sec:motivation} provides the motivation behind the proposed ECC. Section~\ref{blob_scanning} describes blob scanning and registration. The proposed context adaptive chain coding method is presented in Section~\ref{Context-based Chain Coding}, followed by a novel skip coding method in Section~\ref{skip_coding_mode}. The decoding process, which reconstructs the semantic map from the coded bitstream, is described in Section~\ref{Decoding_Process}. To achieve better coding performance, we employ an adaptive arithmetic coder to encode all the syntax elements.
	
	\subsection{Motivation}
	\label{sec:motivation}
	\textit{Extended} source codes are source codes applied to groups of symbols, rather than individual symbols. For example, second extension of a Huffman code~\cite{Sayood} would assign codewords to groups of two symbols, rather than individual symbols. This can bring the rate closer to the entropy of the source. To see this, let $c(a_i)$ be the codeword assigned to symbol $a_i$ and let $l(c(a_i))$ be its length. In well-designed source codes, the codeword length in bits is approximately equal to the negative log-probability of the symbol~\cite{Cover}:
	\begin{equation}
		l(c(a_i)) \approx -\log_2 p(a_i).
	\end{equation}
	When two symbols, say $a_i$ and $a_j$, are encoded, the number of bits spent on them is approximately
	\begin{equation}
		\begin{aligned}
			l(c(a_i)) + l(c(a_j)) 
			& \approx -\log_2 p(a_i) -\log_2 p(a_j) \\
			& = -\log_2 \left( p(a_i)\cdot p(a_j)\right).
		\end{aligned}
	\end{equation}
	This number of bits would be appropriate if the symbols were independent, i.e., if $p(a_i, a_j)=p(a_i)\cdotp p(a_j)$. Otherwise, by the Wrong Code Theorem~\cite{Cover}, the compression performance degrades.
	
	Extended source codes alleviate this issue by encoding groups of symbols based on their joint probability. Thus, a second extension of a code assigns the codeword length to the pair of symbols $(a_i,a_j)$ according to their joint probability,
	\begin{equation}
		l(c(a_i, a_j)) \approx -\log_2 p(a_i, a_j),
	\end{equation}
	thereby providing the number of bits that is more appropriate to the true statistics of the source.
	Despite their benefits, extended source codes face their own challenges, the main ones being: 1) the difficulty of estimating joint probabilities of many symbols and 2) the exponential expansion of the joint codebook: if there are $K$ symbols, there are $K^n$ different blocks of length $n$, and each one would require its own codeword. While there is no easy solution to the former challenge (also known as the \textit{Curse of Dimensionality}), the latter challenge can sometimes be mitigated if the nature of the source makes certain combinations of symbols impossible.
	
	The extended chain code (ECC) we propose in this work can be considered an extension of the F8 code discussed earlier. In particular, the proposed ECC groups 2\textendash3 symbols from F8 in order to exploit their joint probabilities and provide bit savings. Under such conditions, the codebook size could be expected to be between $8^2=64$ and $8^3=512$; however, we end up with only 36 symbols due to the fact that some combinations of symbols from F8 are impossible, such as 1 followed by 5, or 3 followed by 7. Hence, the codebook remains manageable while exploiting the benefits of block encoding. Details of the proposed ECC are presented in the following sections.
	
	\subsection{Blob scanning and registration}
	\label{blob_scanning}
	
	\begin{figure}[!t]
		\centering
		\begin{minipage}[b]{0.41\linewidth}
			\centering
			\includegraphics[width=\textwidth]{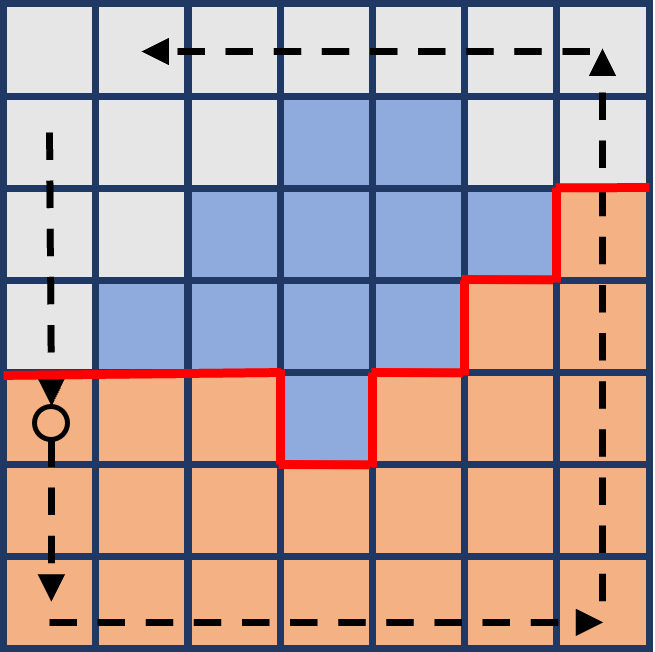}
			\centerline{(a) Boundary blob scan}\medskip
		\end{minipage}
		\hspace{0.05\linewidth}
		\centering
		\begin{minipage}[b]{0.41\linewidth}
			\centering
			\includegraphics[width=\textwidth]{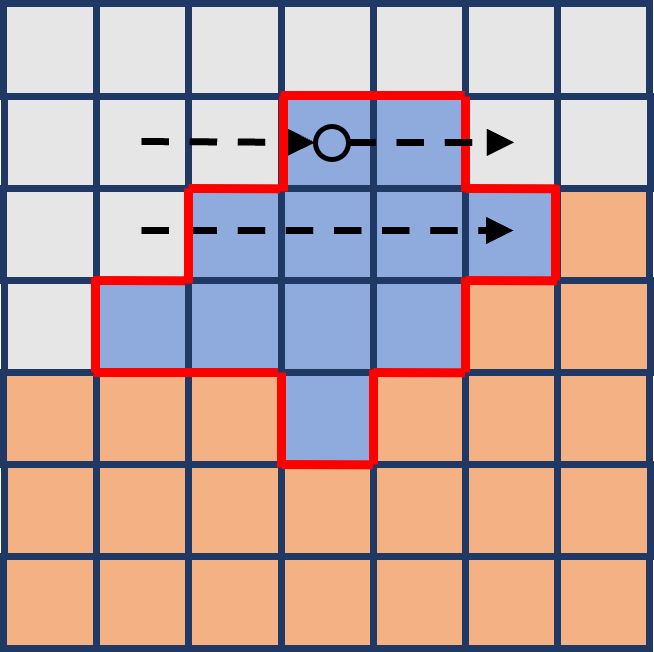}
			\centerline{(b) Inner blob scan}\medskip
		\end{minipage}
		\caption{Scanning directions for (a) boundary blobs and (b) inner blobs, where the red curve indicates the contour of the blob and the circle is recorded as the position of the blob.}
		\label{fig:scanning}
	\end{figure}
	
	The encoding process begins by scanning the input $X$ to find all semantic regions, referred to as \emph{blobs}. A blob is defined as a group of continuously connected pixels associated with the same label. There are two types of blobs in the proposed framework: boundary blobs that touch the frame boundary, and inner blobs that are fully enclosed by other blobs. In the following description, the term \emph{contour} is used also to denote the outline of a blob, which is represented using chain coding for efficient compression. As illustrated in Fig.~\ref{fig:scanning}, the scanning starts at the top-left corner and proceeds along the frame boundary in a counterclockwise direction to identify boundary blobs, followed by a raster-scan pass to identify inner blobs. During scanning, shared portions of the perimeter between adjacent blobs are identified by comparing the labels of neighboring pixels. When a label change is detected, the corresponding perimeter segment is associated with the contour of the current blob. The red curves in Fig.~\ref{fig:scanning} represent the contours of the registered blobs according to the scanning order. The detected point with a label change, indicated by a circle, is recorded as the position of the blob. The numbers of registered boundary blobs and inner blobs, the position of each blob, and the corresponding label required for reconstruction are encoded in the bitstream as metadata. For the top-left boundary blob, only the associated label is coded since the contour of the blob can be automatically derived after the rest of blobs are fully reconstructed in the decoder. 
	
	To encode the position of each blob, we use adaptive Golomb coding to encode differences in scan order instead of Cartesian coordinates (e.g., x and y position). 
	
	To encode the label of each blob, a label palette is built according to the raster-scan order, and the label palette is used to replace the label with the label index at the beginning of the encoding. The label is restored at the decoder using the label palette.
	
	\subsection{Context Adaptive Chain Coding Methods}
	\label{Context-based Chain Coding}
	
	After blob scanning and registration, the contour of each blob is encoded using chain coding. This subsection describes the proposed chain coding methods employed in this compression framework. A novel extended chain code (ECC) is introduced to efficiently represent long contour transitions, while a legacy three-orthogonal chain code (3OT) is used as a fallback mode to improve coding efficiency. By comparing the total number of bits required for each representation, the encoder selects the mode that results in the lowest bitrate.
	
\begin{figure}[!t]
	\centering
	\begin{minipage}[b]{0.5\linewidth}
		\centering
		\includegraphics[width=\textwidth]{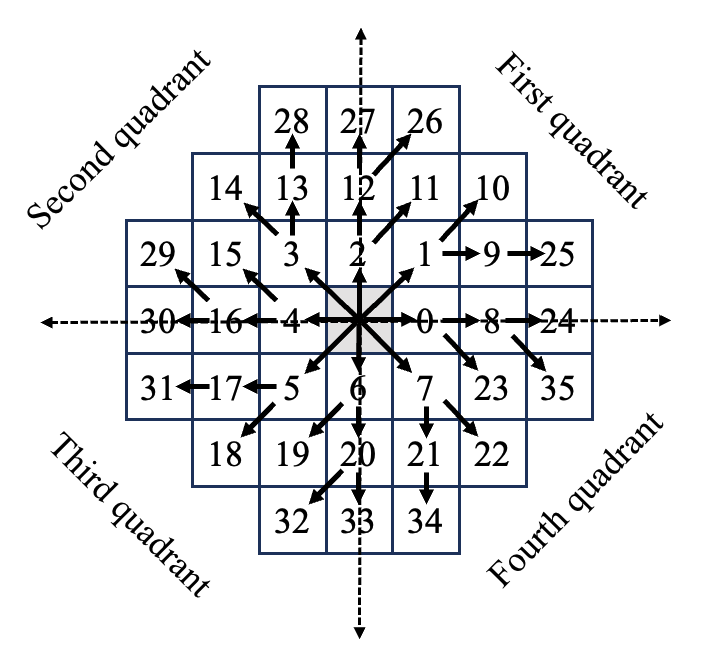}
		\centerline{(a) Extended Chain Code}\medskip
	\end{minipage}
	\hspace{-0.05\linewidth}
	\centering
	\begin{minipage}[b]{0.5\linewidth}
		\centering
		\includegraphics[width=\textwidth]{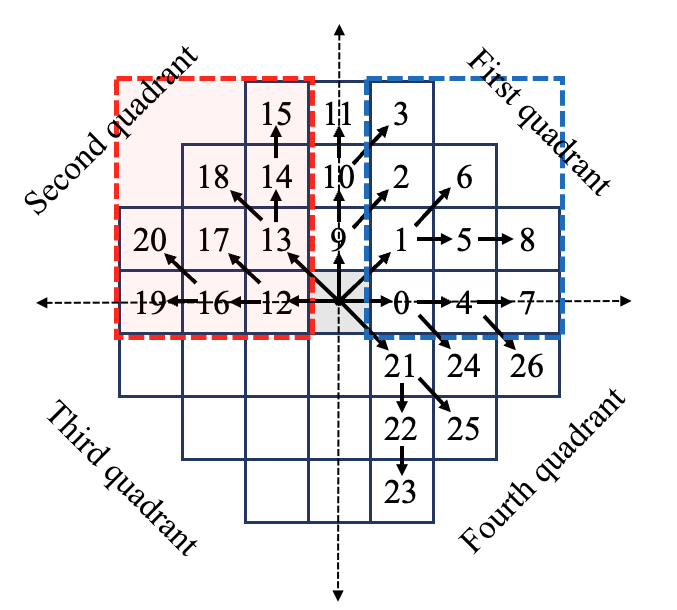}
		\centerline{(b) Relative ECC}\medskip
	\end{minipage}
	\caption{Illustration of the proposed (a) extended chain code (ECC) and (b) Relative ECC (RECC), where the red box highlights symbols with less frequency and the blue box highlights the quadrant to which self-reference symbols are mapped for second-order context modeling.}
	\label{fig:proposed_chain_code}
\end{figure}

	The proposed ECC is designed to encode longer contour transitions using an expanded symbol set, thereby reducing the number of symbols required to represent extended or smooth boundaries. The ECC is constructed by extending conventional Freeman eight-direction (F8) chain codes to cover a larger spatial neighborhood. As shown in Fig.~\ref{fig:proposed_chain_code}(a), the ECC comprises at most three layers of F8 directions, resulting in a total of 36 symbols. For coding efficiency, symbols corresponding to longer spatial displacements are preferred over combinations of multiple shorter displacements. For example, symbol 24 is selected instead of a sequence of symbols 0 and 8 when representing the same contour path. As a result, ECC with an expanded symbol set can represent a given contour using fewer symbols than conventional F4 or F8 chain codes.
	
	Naively encoding a large symbol alphabet, however, increases the number of bits required per symbol. To address this issue, the proposed framework employs ECC as a first-order chain code only for establishing the initial contour direction. Subsequent contour transitions are encoded using a second-order representation referred to as the relative extended chain code (RECC), which serves as a chain-difference representation of ECC symbols. The RECC shown in Fig.~\ref{fig:proposed_chain_code}(b) consists of 27 symbols. The third quadrant of the RECC symbol layout is empty, since transitions corresponding to direct reversal of the previous contour direction do not occur. 
	
	In the proposed relative representation, a self-reference quadrant is defined to normalize symbol orientation; without loss of generality, the first quadrant is used as the self-reference quadrant as highlighted by the blue box in Fig.~\ref{fig:proposed_chain_code}(b). Let $q_{sr}$ denote the self-reference quadrant index. Given the initial ECC symbol $e_t$, and the quadrant index of the symbol is identified as $q(e_t)\in\{0, 1, 2, 3\}$. The rotation angle required to align $e_t$ with the self-reference quadrant is defined as
	\begin{equation}
	\theta_t=norm((q_{sr}-q(e_t))\times90^\circ),
	\end{equation}
	where $norm(\cdot)$ normalizes the angle to the set $\{0^\circ, +90^\circ, \pm180^\circ, -90^\circ\}$. The rotation angle $\theta_t$ is subsequently applied to the next ECC symbol $e_{t+1}$ to obtain the corresponding RECC symbol $r_{t+1}$ by
	\begin{equation}
		r_{t+1}=R(Rot(e_{t+1},\theta_t)),
	\end{equation}
    where $Rot(\cdot,\theta_t)$ rotates the ECC symbol $e_{t+1}$ by $\theta_t$ and $R(\cdot)$ maps the rotated symbol to the corresponding RECC symbol used as the final encoded symbol. The self-reference is defined as $R(Rot(e_t,\theta_t))$ for context modeling.
	
\begin{figure}[!t]
	\centering
	\begin{minipage}[b]{1\linewidth}
		\centering
		\includegraphics[width=\textwidth]{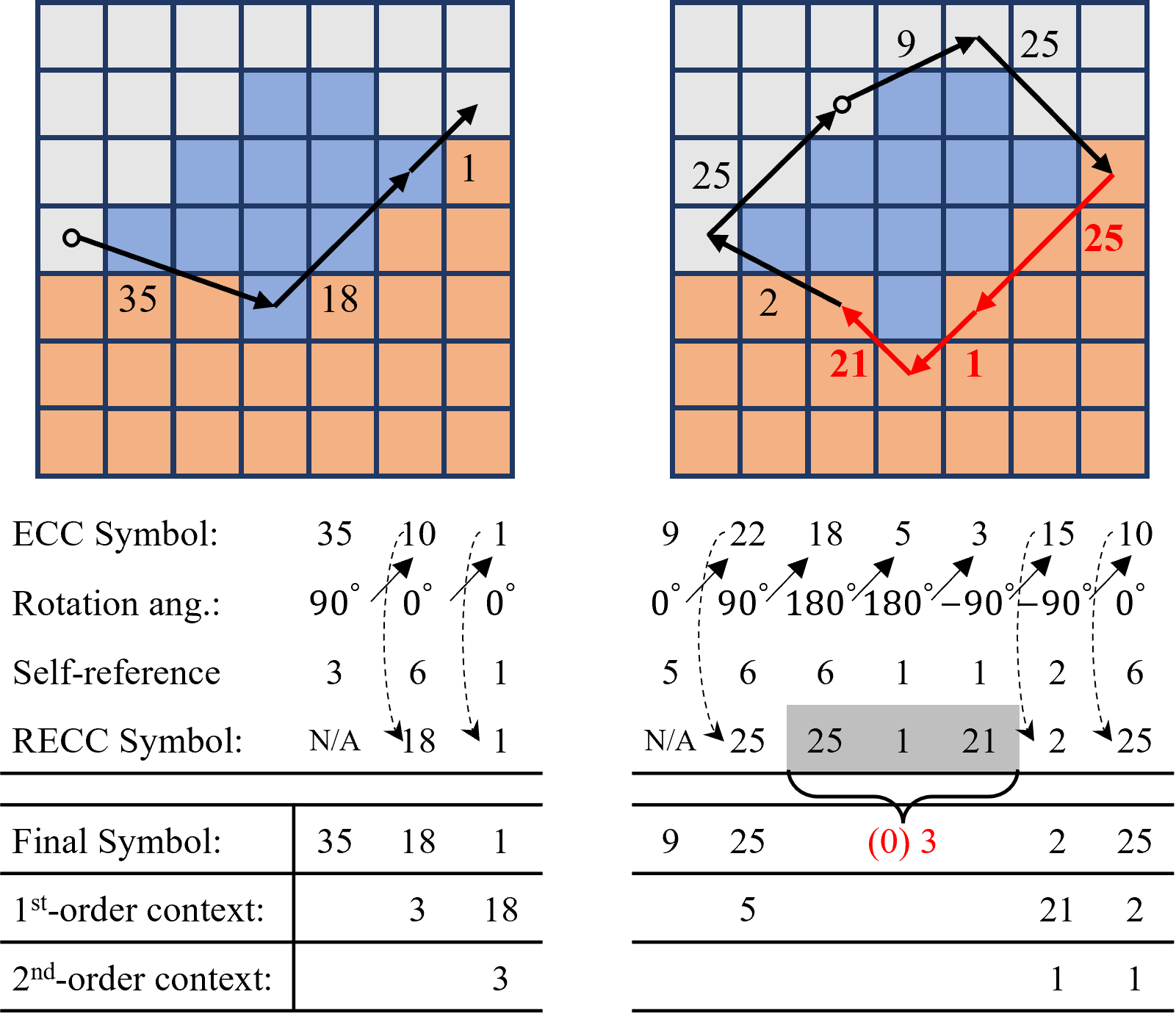}
	\end{minipage}
	\caption{Example of the proposed extended chain coding process for a boundary blob (left) and an inner blob (right), showing ECC symbol assignment, rotation-based mapping to relative ECC (RECC) symbols, self-reference used for context derivation, and the resulting final chain code sequence.}
	\label{fig:chain_code_examples}
\end{figure}

	Fig.~\ref{fig:chain_code_examples} illustrates the proposed chain coding process for representative blobs. The left and right examples present the processing steps for a boundary blob and an inner blob, respectively, organized according to the rows shown at the bottom of the figure. Along the contour starting from the position marked by a circle next to the blob position, an ECC symbol is assigned to each directional edge. During ECC symbol assignment, the corresponding rotation angles are recorded for subsequent derivation of both the RECC symbols and the self-reference used for context modeling. Finally, the contour ends at the map boundary for the boundary blob and ends at the contour starting position for the inner blob.
	
	The first symbol of the final chain code sequence is encoded directly using an ECC symbol. For each subsequent symbol, the current ECC symbol is rotated using the rotation angle recorded for the previously encoded symbol to obtain the final RECC symbol used for encoding. When encoding the second symbol, since no RECC symbol exists for the previous symbol, the self-reference is used as the first-order context. For all remaining symbols, the first-order context is derived from the RECC symbol of the immediately preceding symbol. The second-order context is derived from the self-reference symbol corresponding to the second previous symbol whenever available.	As a result, the maximum number of context tables for encoding each final symbol is $27 \times 9 = 243$. 
	
	In addition, it is observed that the RECC symbols from 12 to 20 occur with relatively low frequency, as highlighted by the red box in Fig.~\ref{fig:proposed_chain_code}(b). To account for this behavior, 9 RECC available flags are introduced to signal the decoder whether the symbol is available, reducing the bitrate. 
	
	In addition to ECC, the proposed framework also evaluates contour encoding using the three-orthogonal chain code (3OT) \cite{cruz2005compressing}, one of the most compact conventional chain codes, as illustrated in Fig.~\ref{fig:F84VCC3OT}. The 3OT chain code consists of three symbols: symbol 0 indicates that the contour direction is preserved, symbol 1 indicates a change in direction that differs from the previous change, and symbol 2 is used otherwise. Given the position of a blob, the initial direction is determined, and the reference turning direction is initialized based on the blob type. Specifically, the previous change is initialized as a left turn for boundary blobs and as a right turn for inner blobs, ensuring consistent interpretation of subsequent direction changes.
	
	To encode symbol sequences using the 3OT chain code, a fourth-order context model is employed, based on the preliminary comparative evaluation presented in Table~\ref{tbl:shapes_result}. For each blob, both ECC-based and 3OT-based representations are evaluated, and the encoder selects the representation that yields the lower total number of bits. The selected coding mode is signaled in the bitstream using a mode flag.

	\begin{figure}[!t]
		\centering
		\begin{minipage}[b]{1\linewidth}
			\centering
			\includegraphics[width=\textwidth]{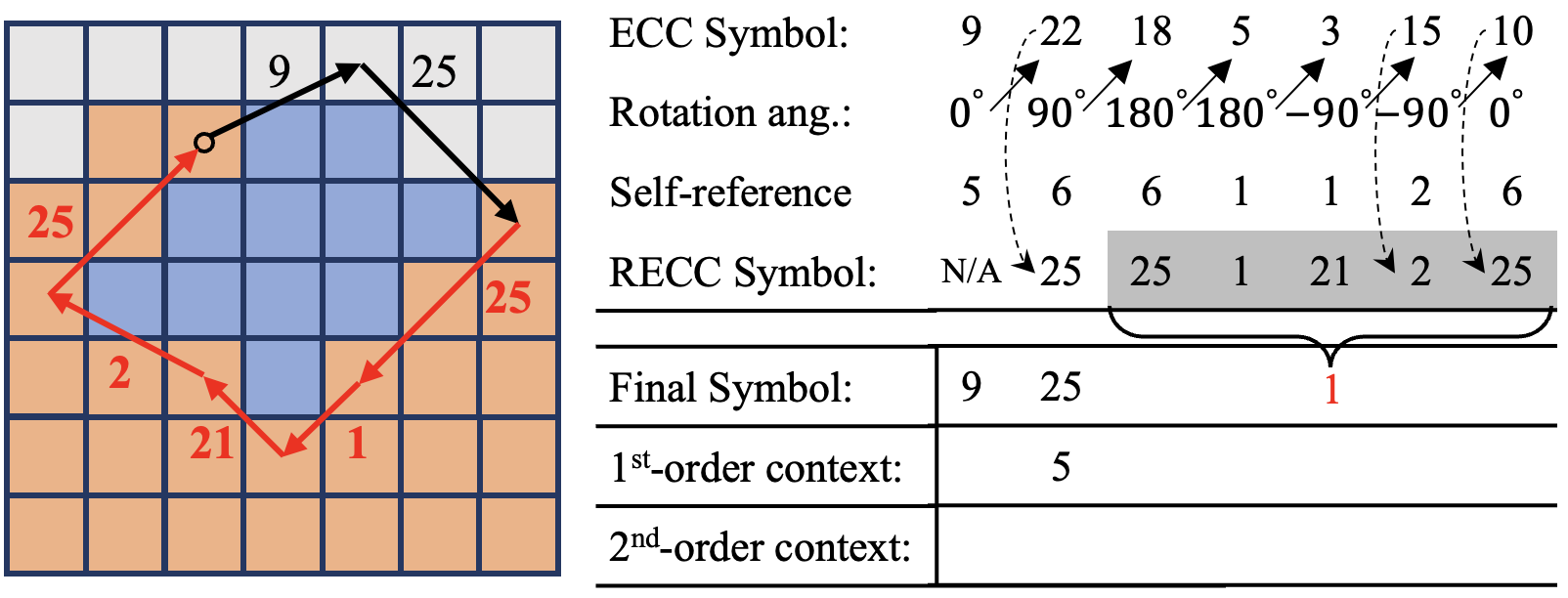}
		\end{minipage}
		\caption{Example of the complete skip mode, in which a subsequence of RECC symbols is replaced by a skip flag equal to 1.}
		\label{fig:full_skip_mode}
	\end{figure}
	
	\subsection{Skip coding mode}
	\label{skip_coding_mode}
	Since blobs are encoded sequentially according to the scanning order described in Section~\ref{blob_scanning}, the encoder maintains a contour occupancy map that records perimeter edges belonging to previously reconstructed blobs. When the contour edge of the current blob coincides with an occupied edge in this map, the edge is identified as the start of a shared perimeter segment. Consecutive shared edges are grouped into the shared perimeter segment, and the corresponding contour symbols are omitted from explicit transmission using the skip coding mode because the decoder can reconstruct the segment from previously decoded blobs. As shown on the left in Fig.~\ref{fig:chain_code_examples}, after the boundary blob is encoded, the inner blob is encoded sequentially. Since the neighboring blob has already been encoded, shared perimeter segments can be readily identified as illustrated by the red directional edges and corresponding symbols on the right in Fig.~\ref{fig:chain_code_examples}. 
	
	To signal the use of skip coding, a \emph{complete skip mode} is introduced. When a sequence of RECC symbols corresponds to a shared perimeter segment that starts and ends within the current blob, the shared perimeter can be inferred at the decoder without explicitly transmitting the corresponding chain code symbols. In such cases, skip coding is applied to reduce the bitrate. If the shared perimeter segment terminates before the end of the contour, skip coding is applied in partial skip mode, where the \emph{complete skip mode} is set to 0 and followed by a run-length value indicating the number of skipped symbols. The run-length value is encoded by the third-order	Exp-Golomb algorithm. In the example shown in Fig.~\ref{fig:chain_code_examples}, three symbols are skipped before normal chain coding resumes. Meanwhile, when the shared perimeter segment extends to the end of the contour, as shown in Fig.~\ref{fig:full_skip_mode}, the \emph{complete skip mode} is activated by setting the flag to 1, and no run-length value is transmitted. By signaling one flag, the remaining contour should be fully inferred at the decoder.
	
	In addition, if the end of the contour is not attached to the shared perimeter segment, there is no need to transmit the \emph{complete skip mode}, since the \emph{complete skip mode} flag must be 0. This mechanism applies to both boundary blobs and the inner blob shown on the right in Fig.~\ref{fig:chain_code_examples}.
	
	\begin{figure}[!t]
		\centering
		\begin{minipage}[b]{0.7\linewidth}
			\centering
			\includegraphics[width=\textwidth]{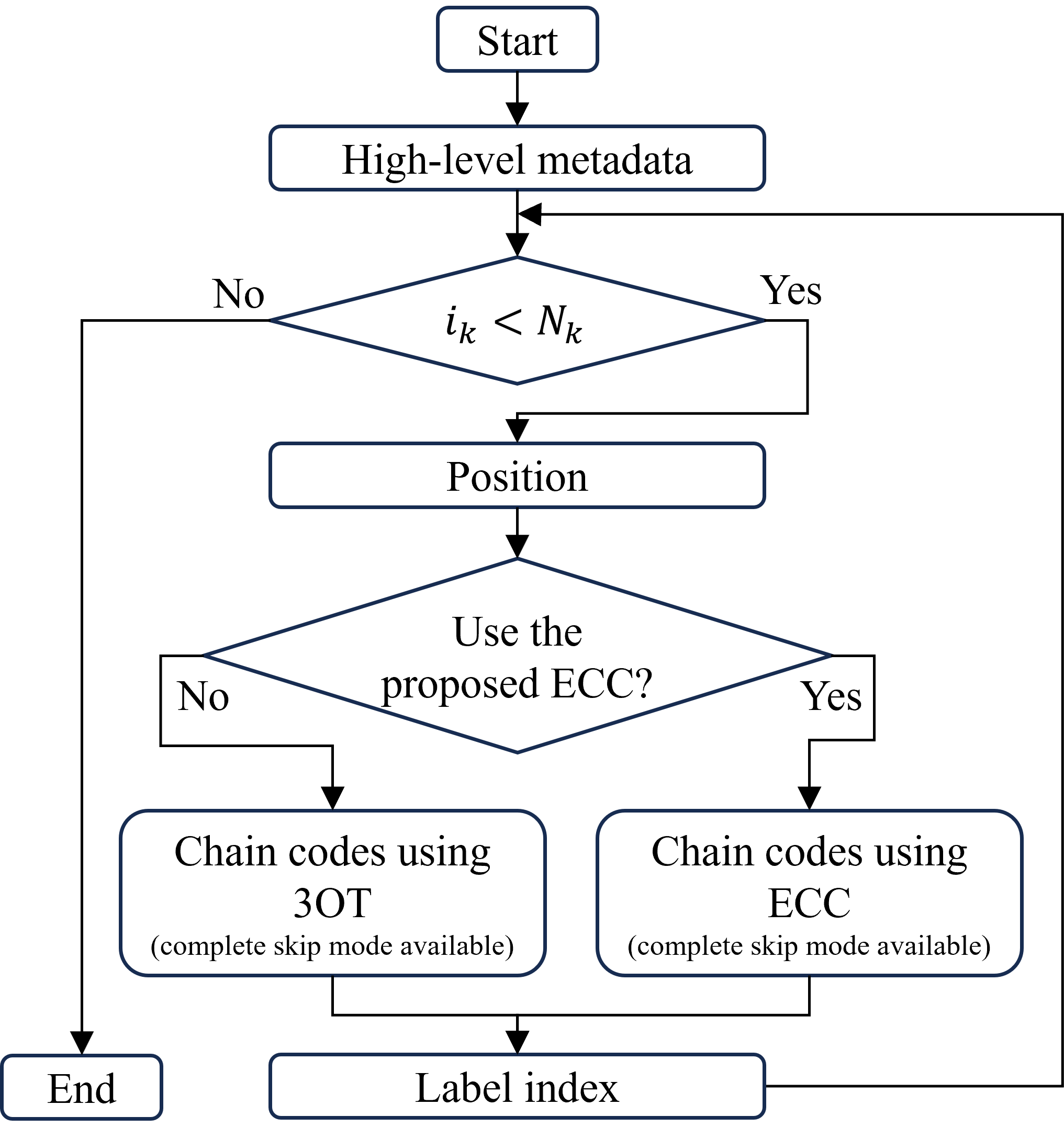}
		\end{minipage}
		\caption{Overview of decoding process of the proposed lossless semantic map compression.}
		\label{fig:decoding_process}
	\end{figure}
	
	\subsection{Decoding Process}
	\label{Decoding_Process}
	
	Since the bitstream is required to be self-decodable, it contains not only chain code symbols, label indexes, and position of each blob but also the necessary high-level metadata for reconstruction. Such metadata includes the size of the input semantic map, the numbers of boundary and inner blobs, the label value assigned to each label index, and RECC available flags. 
	
	The overall decoding procedure is illustrated in Fig.~\ref{fig:decoding_process}. At the beginning of decoding, the high-level parameters are parsed from the bitstream to obtain, for example, the total number of coded blobs $N_k$, where $k \in \{\text{boundary blob}, \text{inner blob}\}$. A counter $i_k$ is maintained to track the number of decoded blobs of each type. The decoding process continues until all blobs have been reconstructed, i.e., until $i_k = N_k$ for each blob type.
	
	\begin{figure*}[ht!]
		\centering
		\begin{minipage}[b]{0.9\linewidth}
			\centering
			\centerline{\includegraphics[width=\linewidth]{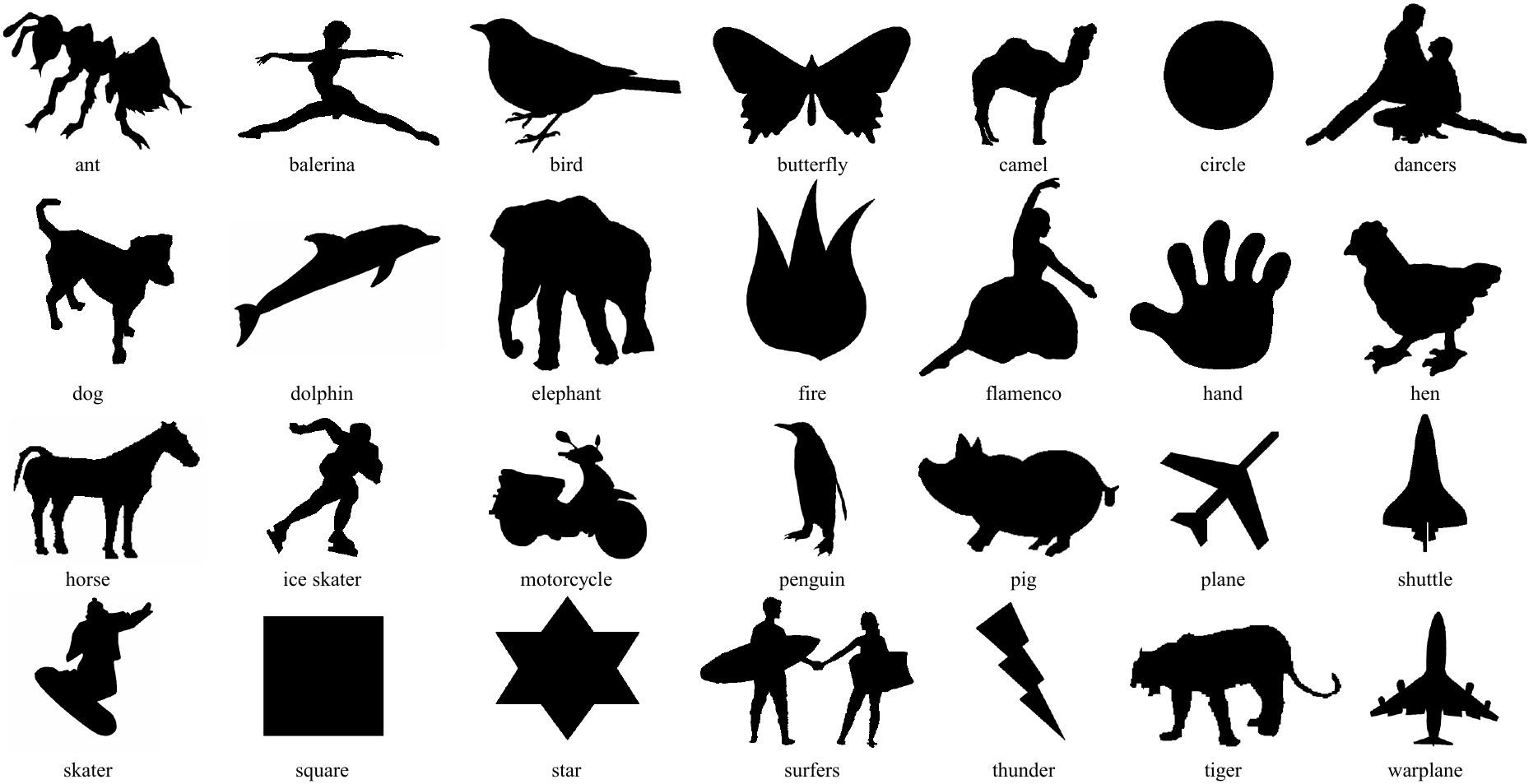}}
		\end{minipage}
		\caption{Visualization of test image shapes used to evaluate the performance of the proposed ECC compared with other methods.}	
		\label{fig:shapes}
	\end{figure*}
	
	To decode an individual blob, the decoder first parses the position of the blob and obtains the starting position of the contour, followed by a chain coding mode flag. If the flag indicates the use of the proposed ECC method, the decoder parses the chain code symbols and reconstructs the blob contour using the context-adaptive chain coding process. Otherwise, when the flag indicates the fallback mode, the decoder parses the chain code symbols using the other chain coding method, three-orthogonal chain code (3OT) based context coding. For both cases, the complete skip mode is available. Finally, the decoder parses the label index of the blob, removing label index candidates attached to the contour to save bitrate.
	
	After decoding the contour, the corresponding semantic region is reconstructed and the associated label information is applied. This procedure is repeated for each blob until the complete semantic map is reconstructed identically to the original input.
	
	\section{Experimental Results}
	\label{sec:experiment}
	This section evaluates the performance of the proposed lossless semantic map compression framework through a series of experiments, comparing it with state-of-the-art chain coding methods and widely used lossless image and semantic map compression methods. The evaluation focuses on compression efficiency and runtime performance, as well as the impact of key components of the proposed framework through ablation studies. All experiments were conducted on a single PC running a 64-bit Windows 11 operating system with an AMD Ryzen 7 5800H processor and 16 GB of memory.
	
	\subsection{Benchmarks}
	\label{benchmarks}
	
	As a preliminary evaluation of the proposed chain coding design, we compare ECC with F8, F4, VCC, and 3OT by counting the total number of symbols required to represent a set of test image shapes obtained from an online source\footnote{\url{https://gemma.feri.um.si//files/Images_V_0.3.7z}}, shown in Fig.~\ref{fig:shapes}. In addition, the effectiveness of the proposed Markov model-based chain coding method using RECC is also  evaluated and compared with several alternatives, including relative F8 (RF8) \cite{liu2005efficient}, VCC, 3OT, and a Burrows--Wheeler transform-based method \cite{vzalik2021lossless}.
	
	For lossless semantic map compression, the proposed framework is compared with several benchmark methods, including the HEVC screen content coding (SCC) reference software SCM-7.0 \cite{xu2015overview}, the Joint Bi-level Image Experts Group (JBIG1) lossless image compression standard \cite{international1993progressive}, the high-performance generic lossless image format (FLIF) \cite{sneyers2016flif}, and the recent chain coding-based semantic map compression framework CC-SMC \cite{yang2024cc}, using 3OT-based three-order context coding as the chain coding method. Most recent learning-based compression methods are lossy image compression methods. We did not identify a recent learning-based compression method that focuses on lossless semantic map compression.
	
	To enable palette mode in SCM-7.0, which is a specialized coding tool for lossless compression of screen content, each semantic map is converted to YUV420 format with the U and V channels set to zero prior to encoding. For JBIG1, preprocessing is required since the codec is designed for bi-level image coding. Specifically, a mapping table is constructed to convert multi-level semantic values into indices, and the semantic map is decomposed into multiple bit planes, each of which is encoded using JBIG1. At the decoder, the bit planes are recomposed and mapped back to the original semantic values. 
	
	\begin{table}
		\caption{Comparison of various chain code designs in terms of the number of symbols required to represent a set of geometric shapes. Bold numbers indicate the least number of symbols.}
		\begin{center}
			\begin{tabular}{c|ccccc}
				\hline
				Shape & F8 & F4 & VCC & 3OT & ECC\\
				\hline
				ant  & 4720 & 6494 & 6498 & 6498 & \textbf{2169}\\
				balerina & 1084 & 1440 & 1444 & 1444 & \textbf{485}\\
				bird & 1725 & 2354 & 2358 & 2358 & \textbf{751}\\
				butterfly & 1375 & 1996 & 2000 & 2000 & \textbf{669}\\
				camel & 1324 & 1782 & 1786 & 1786 & \textbf{588}\\
				circle & 758 & 1068 & 1072 & 1072 & \textbf{346}\\
				dancers & 2915 & 4156 & 4160 & 4160 & \textbf{1506}\\
				dog & 1145 & 1546 & 1550 & 1550 & \textbf{495}\\
				dolphin & 2007 & 2870 & 2874 & 2874 & \textbf{942}\\
				elephant & 3410 & 4502 & 4506 & 4506 & \textbf{1485}\\
				fire & 2138 & 2908 & 2912 & 2912 & \textbf{957}\\
				flamenco & 2321 & 3372 & 3376 & 3376 & \textbf{1105}\\
				hand & 1372 & 1858 & 1862 & 1862 & \textbf{599}\\
				hen & 2466 & 3404 & 3408 & 3408 & \textbf{1120}\\
				horse & 3781 & 4310 & 4314 & 4314 & \textbf{1557}\\
				ice skater & 993 & 1432 & 1436 & 1436 & \textbf{483}\\
				motorcycle & 3670 & 5012 & 5016 & 5016 & \textbf{1627}\\
				penguin & 1510 & 1968 & 1972 & 1972 & \textbf{654}\\
				pig & 733 & 1046 & 1050 & 1050 & \textbf{345}\\
				plane & 1490 & 2258 & 2262 & 2262 & \textbf{674}\\
				shuttle & 971 & 1230 & 1234 & 1234 & \textbf{394}\\
				skater & 1510 & 2184 & 2188 & 2188 & \textbf{707}\\
				square & 1092 & 1092 & 1096 & 1096 & \textbf{366}\\
				star & 1024 & 1444 & 1448 & 1448 & \textbf{497}\\
				surfers & 1973 & 2646 & 2650 & 2650 & \textbf{904}\\
				thunder & 1442 & 2476 & 2480 & 2480 & \textbf{759}\\
				tiger & 3703 & 4250 & 4254 & 4254 & \textbf{1485}\\
				warplane & 2019 & 2658 & 2662 & 2662 & \textbf{881}\\
				\hline
				Total & 54671 & 73756 & 73868 & 73868 & \textbf{24550} \\
				\hline
			\end{tabular}
			\label{tbl:random_shapes}
		\end{center}
	\end{table}
	
	\begin{table}
		\caption{Comparison of CAECC (Ours) with the other chain coding methods in terms of bits, with the best results in bold. Subscript value shows the maximum number of context order.}
		\begin{center}
			\begin{tabular}{c|cccccc}
				\hline
				Shape & RF8${_4}$ & VCC${_5}$ & 3OT${_4}$ &C$^*$& \cite{vzalik2021lossless} & CAECC${_2}$\\
				\hline
				horse      & 4152 & 2911 & 3071 &4881& \textbf{2400} & 3954 \\
				ice skater & 1311 & 1430 & 1377 &1714& 1536 & \textbf{1255} \\
				motorcycle & 3784 & 4079 & 4034 &5430& 4324 & \textbf{3679} \\
				penguin    & 1786 & 1915 & 1869 &2421& 2160 & \textbf{1671} \\
				pig        & 979  & 1110 & 1045 &1293& 1199 & \textbf{953} \\
				plane      & 1161 & 1028 & 1289 &1348& \textbf{828}  & 1140 \\
				shuttle    & 919  & 1025 & 922  &1163& 933  & \textbf{845} \\
				skater     & 1672 & 1883 & 1839 &2293& 2090 & \textbf{1620} \\
				square     & 401  & 72   & 67   &329& \textbf{40}   & 314 \\
				star       & 778  & 732  & 655  &1034& \textbf{385}  & 787 \\
				surfers    & 2633 & 2787 & 2737 &3411& 3156 & \textbf{2597} \\
				thunder    & 1196 & 1316 & 1361 &1084& \textbf{752}  & 1242 \\
				tiger      & 4510 & 3415 & \textbf{3311} &4749& 4003 & 4258 \\
				warplane   & 2463 & 2606 & 2494 &3137& 2951 & \textbf{2348} \\
				\hline
				Total        & 27745 & 26309 & \textbf{26071} &34287& 26757 & 26663 \\
				\hline		
			\end{tabular}
			\label{tbl:shapes_result}
		\end{center}
	\end{table}
	
	\subsection{Preliminary performance of chain coding methods}
	\label{ssec:preliminary}
	
	We first conduct a preliminary evaluation of the proposed ECC by measuring the number of symbols required to represent a set of geometric shapes, in comparison with other chain coding methods. From the original set of 32 shapes, 28 shapes are selected by excluding those containing holes, which result in multiple disjoint contours. The selected shapes, shown in Fig.~\ref{fig:shapes}, span a range of geometric complexity.
	
	As summarized in Table~\ref{tbl:random_shapes}, the total number of ECC symbols required to represent the selected shapes is less than half of that required by the F8 chain code and approximately one third of that required by the other chain code designs. These results demonstrate the representational efficiency of the proposed ECC in terms of symbol count. 
	
	Yet, a reduction in symbol count alone does not directly translate to improved compression efficiency, which motivates the proposed context-adaptive extended chain coding method (CAECC). Other compared chain coding methods also employ context modeling to reduce total bitrate. To initialize the context probability models for RF8 and CAECC, the first 14 shapes (corresponding to the first two rows in Fig.~\ref{fig:shapes}) are used, while the remaining 14 shapes are reserved for evaluation. Whether to initialize the context probability model and the choice of context order depends on the bitrate.
	
	Table~\ref{tbl:shapes_result} summarizes the chain coding performance of CAECC in comparison with RF8, VCC, 3OT, C$^*$ coding \cite{li2026moric} and the method proposed by {\v{Z}}alik~\textit{et al.}~\cite{vzalik2021lossless} in terms of coded bits. Among the evaluated methods, 3OT with a fourth-order context model (3OT$_4$) achieves the lowest total bit count across the tested shapes. One advantage of 3OT is that it can adapt the probability distribution efficiently using a small symbol alphabet and a relatively small context table number. This approach works well for both simple and complex contours, such as the \texttt{square} and \texttt{tiger} shapes. This observation motivates the use of 3OT as a fallback coding mode in the proposed framework.
	
	Nevertheless, the proposed CAECC achieves the lowest bit count on 8 out of the 14 evaluated shapes compared with the other methods. It is particularly suitable for regular contours. It is also noteworthy that CAECC attains competitive compression performance with a limited context order of two, highlighting its efficiency in balancing context complexity and coding performance.
	
	\subsection{Evaluation on Semantic Maps}
	\label{Results of Semantic Map Coding}
	
	\begin{table}
		\caption{Summarized characteristics of tested semantic map datasets.}
		\begin{center}
			\begin{tabular}{c|cccc}
				\hline
				Dataset & Resolution & Maps & Max Values & Avg. \# of Blobs\\
				\hline
				CASIA-B  & $320 \times 240$ & 8443 & 2& 1.16\\
				\hline
				DAVIS 480p & $848 \times 480$ & 6268 & 5& 40.73\\
				\hline
				Cityscapes & $2048 \times 1024$ & 2975 & 27& 123.95\\
				\hline
			\end{tabular}
			\label{tbl:datasets}
		\end{center}
	\end{table}
	
	Finally, the proposed lossless semantic map compression framework is evaluated on three widely used large-scale semantic map datasets: CASIA-B~\cite{yu2006framework}, DAVIS~\cite{perazzi2016benchmark}, and Cityscapes~\cite{cordts2016cityscapes}. These datasets collectively cover a broad spectrum of semantic maps, ranging from simple to complex structures. Detailed characteristics of the test datasets—including resolution, number of maps, maximum value count, and average number of blobs—are summarized in Table~\ref{tbl:datasets}.
	
\begin{figure}
		\centering  
		\subfigure[]{
			\begin{minipage}[b]{0.15625\linewidth}
				\includegraphics[width=\linewidth]{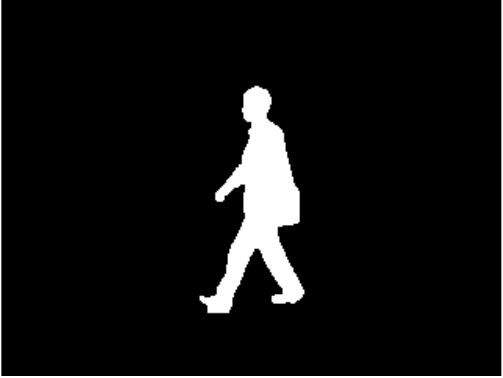}
			\end{minipage}
			\label{fig:semantic1}
		}
		\hspace{0.5cm}
		\subfigure[]{
			\begin{minipage}[b]{0.4140625\linewidth}
				\includegraphics[width=\linewidth]{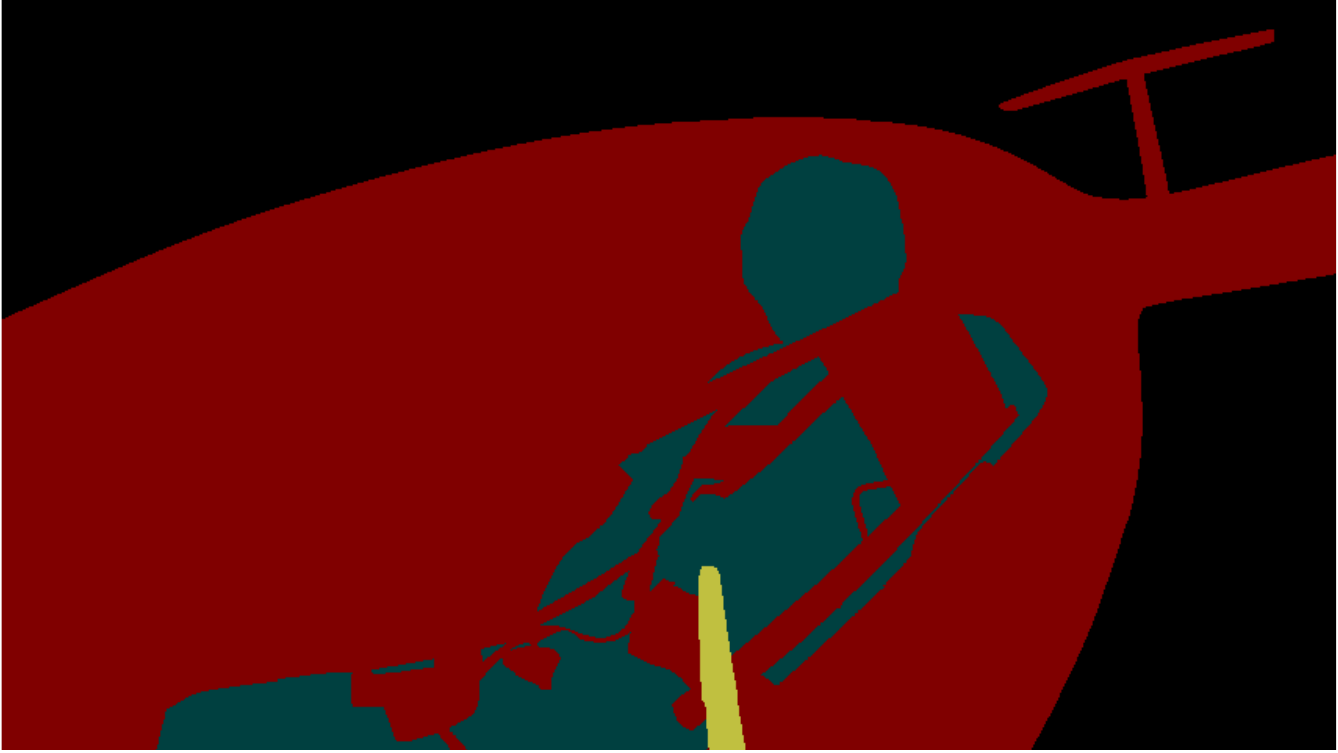}
			\end{minipage}
			\label{fig:semantic2}
		}
		\subfigure[]{
			\begin{minipage}[b]{1.0\linewidth}
				\includegraphics[width=\linewidth]{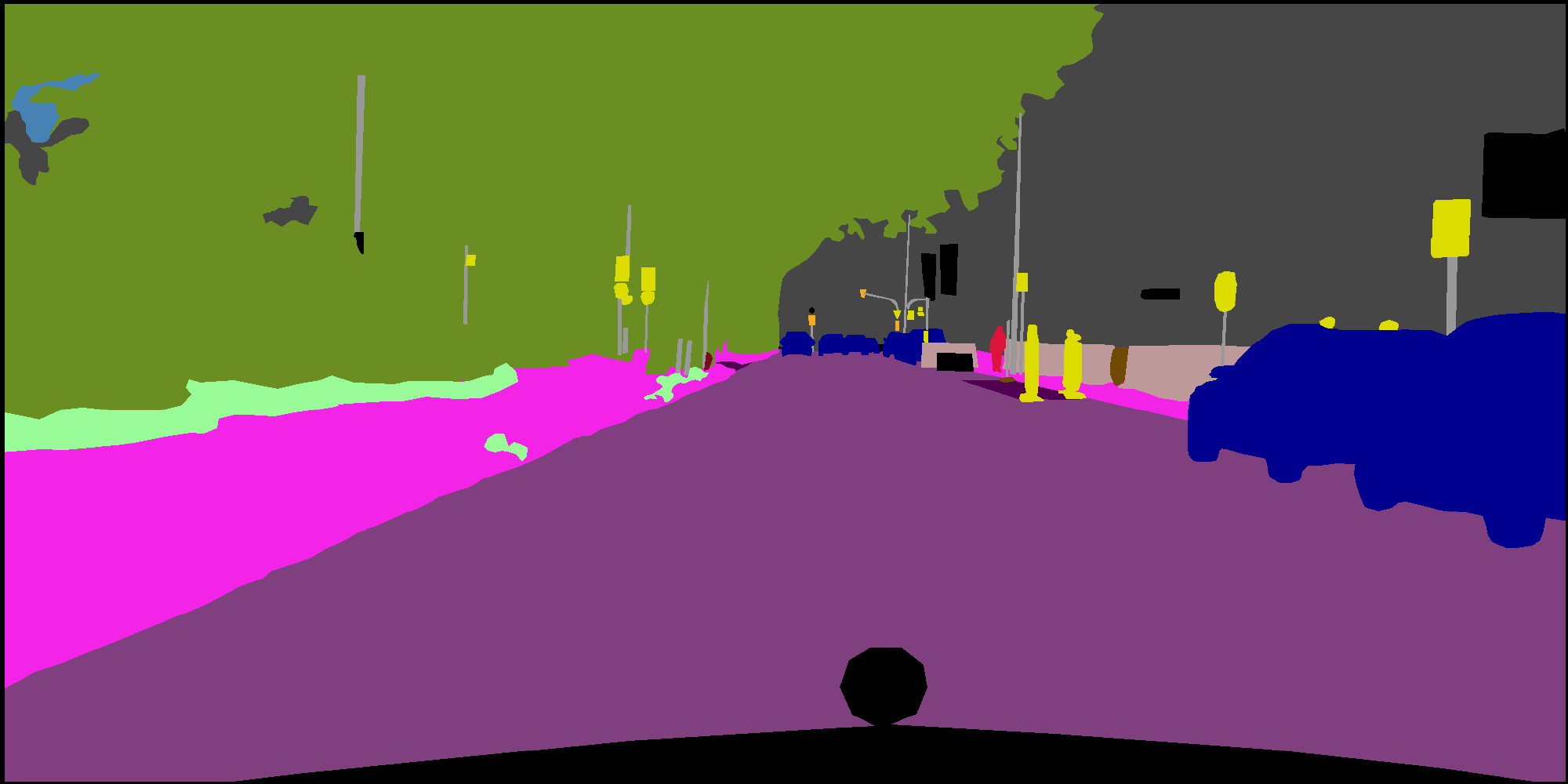}
			\end{minipage}
			\label{fig:semantic3}
		}
		\caption{Sample images from each datasets: (a) CASIA-B, (b) DAVIS 480p, (c) Cityscapes.}
		\label{fig:semantic}
	\end{figure}

	CASIA-B is a public gait recognition dataset containing 124 subjects captured from 11 viewpoints. Due to the large scale of the dataset, the first subject is used for evaluation. Because the main difference in the map lies in the viewpoint, this does not compromise the generality of the results. DAVIS is a densely annotated video segmentation dataset available in both 480p and 1080p resolutions. In this study, the 480p version is used, with an original resolution of $854 \times 480$. To comply with the $8 \times 8$ minimum coding unit (CU) size requirement of SCM-7.0, the resolution is cropped to $848 \times 480$. For Cityscapes, the coding methods are evaluated using the training set. Sample images from each dataset are shown in Fig.~\ref{fig:semantic}.
	
	\begin{table*}[ht]
		\caption{Summary of the averaged performance of various lossless semantic map compression, including JBIG1, FLIF, SCM-7.0, CC-SMC~\cite{yang2024cc}, and the proposed context-adaptive ECC, evaluated on three datasets.}
		\begin{center}
			\begin{tabular}{c|c|c|c|c|c|c|c}
				\hline
				\multicolumn{2}{c|}{Test Dataset}&JBIG1&FLIF&SCM-7.0& CC-SMC~\cite{yang2024cc}& Ours &Over CC-SMC\\
				\hline
				\multirow{3}{*}{CASIA-B}& Rate (Bytes) &187.4&136.7&272.9&75.8& \textbf{64.9}&-14.38\%\\
				& Enc. Time (sec) &0.0223&0.0472&0.3261&\textbf{0.0157}& 0.0181&\\
				& Dec. Time (sec) &0.0025&0.0257&0.0109& \textbf{0.0013} & \textbf{0.0013}&\\
				\hline
				\multirow{3}{*}{DAVIS}& Rate (Bytes) &620.7&618.7&997.5&459.1& \textbf{367.5}&-19.95\%\\
				& Enc. Time (sec) & \textbf{0.0245} &0.1948&1.2831&0.0557&0.0524&\\
				& Dec. Time (sec) &0.0275&0.0341&0.0542&0.0106& \textbf{0.0044} &\\
				\hline
				\multirow{3}{*}{Cityscapes}& Rate (Bytes) &6829.8&4882.8&6314.9&3549.4& \textbf{2661.7}&-25.01\%\\
				& Enc. Time (sec) & \textbf{0.0701} &0.9461&6.6576&0.2003&0.1494&\\
				& Dec. Time (sec) &0.1120&0.1175&0.1472& \textbf{0.0671} &0.0876&\\
				\hline
			\end{tabular}
			\label{tbl:results_on_semantic_maps}
		\end{center}
	\end{table*}
	
	\begin{table*}[ht]
		\centering
		\begin{threeparttable}[b]
			\caption{Performance summary of the ablation study for the proposed compression framework. The number in parentheses in the third column indicates the mode selection percentage for each chain code method.}
			\begin{tabular}{c|c|c|c|c|c|c|c}
				\hline
				\multicolumn{2}{c|}{Test Dataset}& Fully integrated our framework & 3OT only & ECC only & ECC + full RECC &ECC + 1-order&w/o skip coding \\
				\hline
				\multirow{3}{*}{CASIA-B}&Rate (Bytes)&64.9 (ECC: 64.8\% 3OT: 35.2\%)&65.4&65.1&67.0&65.0&64.9\\
				&Enc. (sec)&0.0181&0.0174&0.0164&0.0166&0.0160&0.0186\\
				&Dec. (sec)&0.0013&0.0014&0.0012&0.0012&0.0012&0.0013\\
				\hline
				\multirow{3}{*}{DAVIS}&Rate (Bytes)&367.5 (ECC: 32.1\% 3OT: 67.9\%)&372.7&380.2&384.0&395.3&398.0\\
				&Enc. (sec)&0.0524&0.0480&0.0485&0.0492&0.0487&0.0614\\
				&Dec. (sec)&0.0044&0.0040&0.0043&0.0047&0.0044&0.0054\\
				\hline
				\multirow{3}{*}{Cityscapes}&Rate (Bytes)&2661.7 (ECC: 51.1\% 3OT: 48.9\%)&2742.2&2706.6&2713.6&2815.7&4492.1\\
				&Enc. (sec)&0.1494&0.1360&0.1414&0.1461&0.1453&0.1440\\
				&Dec. (sec)&0.0876&0.0778&0.0884&0.0821&0.0858&0.1070\\
				\hline
			\end{tabular}
			\label{tbl:ablation}
		\end{threeparttable}
	\end{table*}
	
	The evaluated performance of various compression methods on the three datasets is summarized in Table~\ref{tbl:results_on_semantic_maps}. On average, the proposed framework achieves the lowest bitrate across all datasets. In particular, compared with the most recent semantic map compression framework, CC-SMC~\cite{yang2024cc}, the proposed framework reduces bitrate by approximately 20\% on average.
	
	While the proposed framework attains the minimum bitrate, its encoding and decoding runtimes remain competitive with those of other coding methods. On the Cityscapes dataset, the proposed framework exhibits longer decoding time than CC-SMC. This behavior arises because the proposed framework performs global scanning during decoding, whereas CC-SMC partitions the semantic map into coding tree units (CTUs), which reduces overall decoding time. Such global scanning becomes a bottleneck for high-resolution maps (e.g., Cityscapes), but this effect is less pronounced for lower-resolution datasets (e.g., DAVIS at 480p).
	
	The proposed framework has relatively low memory overhead because it is based on traditional contour-based chain coding and context-adaptive entropy coding, rather than on deep neural networks or large learned parameter models. The primary memory requirements arise from storing the contour occupancy map for skip coding, the context probability tables for RECC and 3OT coding, and temporary buffers used during contour traversal. Since the RECC context model uses at most 243 context tables and the 3OT fallback mode employs a compact three-symbol alphabet, the context modeling overhead remains limited. Together with the runtime results reported in Table~\ref{tbl:results_on_semantic_maps}, these characteristics indicate that the proposed framework is feasible for deployment in resource-constrained environments, such as edge devices and embedded systems for semantic map processing.
	
	\subsection{Ablation Study}
	\label{Ablation Study}
	
	We focus on analyzing two adopted chain coding methods: the proposed ECC and the 3OT as a fallback mode. As shown in Section~\ref{ssec:preliminary}, the Markov model-based 3OT method demonstrates competitive coding performance, which motivates its integration into the proposed compression framework. Through the ablation study, the effectiveness of each chain coding method is analyzed, along with the impact of skip coding mode and disabling rarely used symbols (The symbols highlighted by the red box in Fig.~\ref{fig:proposed_chain_code}(b)).
		\begin{figure}[t]
		\centering  
		\subfigure[]{
			\begin{minipage}[b]{0.225\linewidth}
				\includegraphics[width=\linewidth]{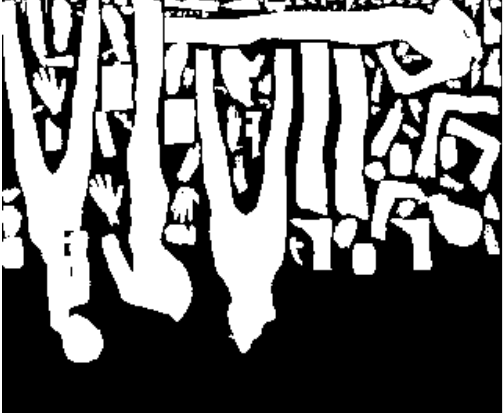}
			\end{minipage}
			\label{fig:occupancy1}
		}
		\subfigure[]{
			\begin{minipage}[b]{0.45\linewidth}
				\includegraphics[width=\linewidth]{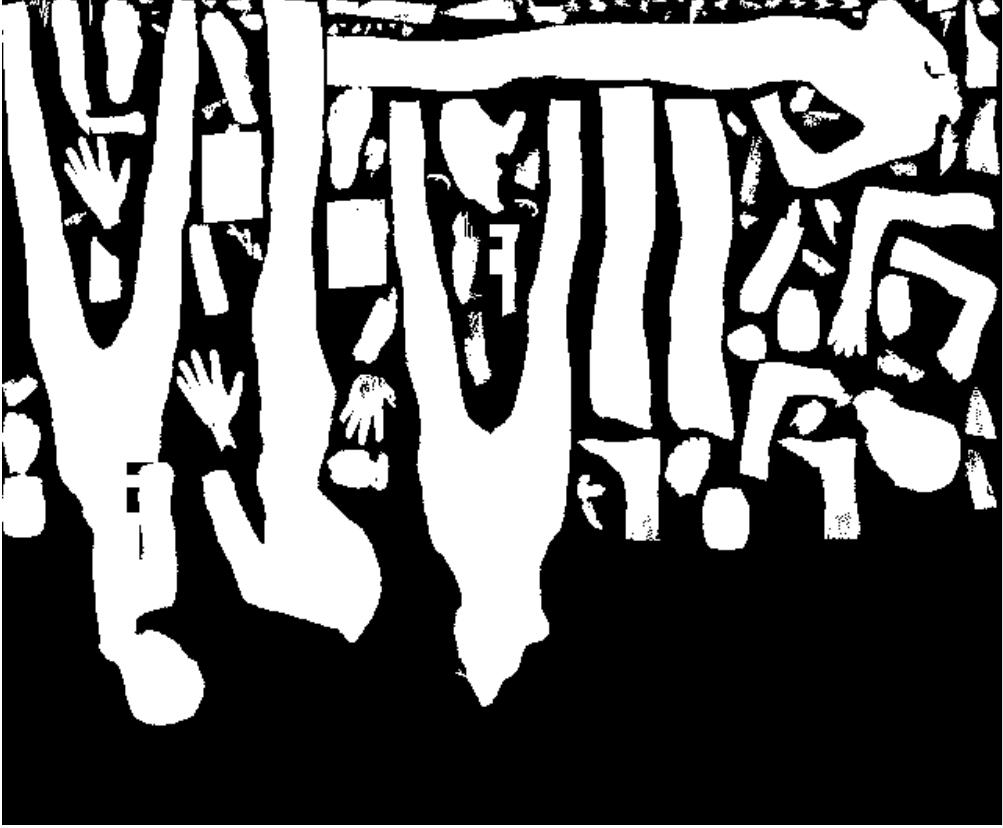}
			\end{minipage}
			\label{fig:occupancy2}
		}
		\subfigure[]{
			\begin{minipage}[b]{0.9\linewidth}
				\includegraphics[width=\linewidth]{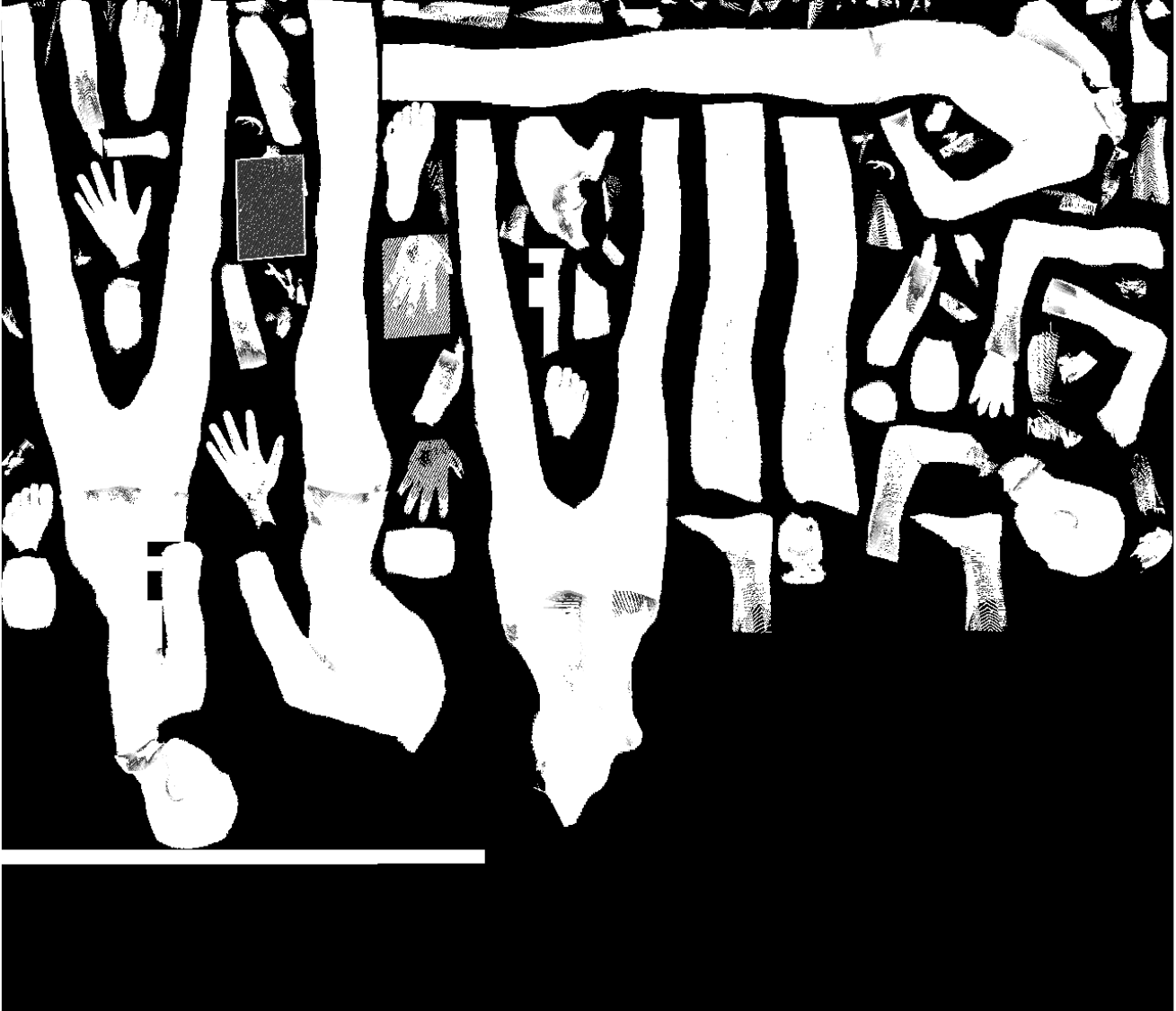}
			\end{minipage}
			\label{fig:occupancy3}
		}
		\caption{Examples of occupancy maps from the V-PCC sequence \texttt{Queen}: (a) $4 \times 4$ downsampled, (b) $2 \times 2$ downsampled, and (c) non-downsampled.}
		\label{fig:occupancy}
	\end{figure}
	\begin{table*}
		\caption{Summary of the averaged performance of various lossless semantic map compression methods evaluated on occupancy maps.}
		\begin{center}
			\begin{tabular}{c|c|c|c|c|c|c|c}
				\hline
				Test Occupancy Map&JBIG1&FLIF&SCM-7.0 Intra&SCM-7.0 Low Delay&CC-SMC Intra&CC-SMC Inter&Ours\\
				\hline
				Loot&182344&172955&256761&254965&176198&175829&\textbf{144602}\\
				Loot $2 \times 2$&66428&68299&106945&96307&59039&58671&\textbf{49807}\\
				Loot $4 \times 4$&33519&34613&50561&44334&30518&29880&\textbf{25492}\\
				RAB&312980&294818&389454&400903&306267&307279&\textbf{287684} \\
				RAB $2 \times 2$&83168&88859&127340&128601&80245&80288&\textbf{71111}\\
				RAB $4 \times 4$&41084&43256&59088&58259&39796&39824&\textbf{35943}\\
				Soldier&346010&331888&476747&456839&340981&338007&\textbf{288079} \\
				Soldier $2 \times 2$&124231&131380&193908&172419&117946&115610&\textbf{104141}\\
				Soldier $4 \times 4$&61327&64866&91571&73219&60188&57250&\textbf{54949}\\
				Queen&572287&538072&614910&\textbf{424195}&573880&478501&609293\\
				Queen $2 \times 2$&106949&115041&152031&112087&104542&\textbf{92114}&94757 \\
				Queen $4 \times 4$&46881&49706&65837&52647&46790&\textbf{40937}&41581 \\
				LD&218578&207006&281819&287492&211599&211998&\textbf{189964} \\
				LD $2 \times 2$&70942&74276&109527&107996&68517&68501&\textbf{59386}\\
				LD $4 \times 4$&35103&36596&50924&49824&33214&33226&\textbf{28885}\\
				\hline
				Rate (Bytes)&153454&150109&201828&181339&149981&141861&\textbf{139045}\\
				\hline
				Enc. Time (sec)&0.34&11.89&83.28&522.65&2.09&20.9&1.16\\
				\hline
				Dec. Time (sec)&0.21&1.84&1.45&1.34&0.22&0.22&0.31\\
				\hline
			\end{tabular}
			\label{tbl:occupancy_map}
		\end{center}
	\end{table*}
	
	Table~\ref{tbl:ablation} summarizes the performance results of the ablation study. The results indicate that ECC is selected more frequently than 3OT on the CASIA-B and Cityscapes datasets, which contain larger and more regular contours. In contrast, for the DAVIS dataset, which includes more irregular contours such as small and fragmented blobs, 3OT is more frequently selected as the optimal chain coding mode. These trends are reflected in the rate performance shown in the fourth and fifth columns of Table~\ref{tbl:ablation}. By combining ECC and 3OT, the proposed framework effectively adapts to diverse contour characteristics and achieves improved overall coding performance.
	
	The sixth column of Table~\ref{tbl:ablation} highlights the impact of the flag controlling the use of the RECC. When all RECC symbols are enabled, the bitrate increases across all datasets, indicating overhead introduced by rarely used symbols. This result demonstrates the importance of selectively disabling such symbols to improve coding efficiency.
	
	The seventh column of Table~\ref{tbl:ablation} highlights the impact of different context orders for coding RECC. The second-order context model improves coding efficiency compared with first-order context model on datasets with long contours and they have similar performance on CASIA-B dataset with short contours. The reason is that complex context model requires more symbols to build accurate probability distribution.
	
	Finally, the last column of Table~\ref{tbl:ablation} shows the impact of the skip coding mode exploiting shared contours. It does not change the performance of binary semantic maps like CASIA-B dataset, but significantly improves the performance of multi valued semantic maps.
	
	\subsection{Extended evaluation on Occupancy Maps}
	\label{Results of Occupancy Map Coding}
	
	Although the proposed framework is designed for lossless semantic map compression, it is also applicable to occupancy maps due to their similar structural characteristics. To investigate this capability, the experimental evaluation is extended to occupancy map compression and compared with other coding approaches. The evaluation is conducted on five dynamic point clouds (DPCs) specified in the MPEG video-based point cloud compression (V-PCC) common test conditions~\cite{schwarz2018common}. For each DPC, dynamic occupancy maps are extracted across 32 frames.
	
	The occupancy maps are evaluated at three spatial resolutions: original resolution (non-downsampled), $2 \times 2$ downsampled, and $4 \times 4$ downsampled. Examples from the V-PCC DPC sequence \texttt{Queen} at the three resolutions are shown in Fig.~\ref{fig:occupancy}. For performance evaluation, SCM-7.0 is configured in intra mode and low-delay mode, and CC-SMC operates in intra mode and inter mode. Table~\ref{tbl:occupancy_map} summarizes the average lossless coding performance of four benchmark methods and the proposed framework.
	
	The proposed framework outperforms the other methods on most occupancy maps, with the exception of the \texttt{Queen} sequence. This behavior can be attributed to the characteristics of the \texttt{Queen} sequence, which contains a large number of fragmented blobs, as illustrated in Fig.~\ref{fig:occupancy}. The inter coding performance of the sequence \texttt{Queen} is much better than the other sequences because	the frame rate of the sequence Queen is 50fps while the	frame rate of other sequences is 30fps. The proposed framework achieves the lowest average bitrate overall, although it does not adopt inter coding tools.	In terms of runtime performance, the proposed framework exhibits much lower encoding time and slightly longer decoding time compared to CC-SMC. 
	
	Overall, since occupancy map bitrate constitutes a significant portion of the total bitrate in lossy V-PCC, employing the proposed framework for occupancy map compression has the potential to improve overall V-PCC coding efficiency.
	
	\section{Conclusion}
	\label{sec:conclusion}
	This paper introduced a contour-based framework for lossless semantic map compression that avoids block-based partitioning and instead exploits contour structure and shared boundaries between adjacent regions. By combining an extended chain code with relative representation and context-adaptive coding, the proposed framework achieves efficient contour description. Experimental evaluation confirms the effectiveness of the proposed design across diverse semantic map scenarios, and further demonstrates its applicability to occupancy maps used in point cloud compression.

	\bibliographystyle{IEEEtran}
	\bibliography{refs}
	
	\begin{IEEEbiography}
		[{\includegraphics[width=1in,height=1.25in, clip,keepaspectratio]{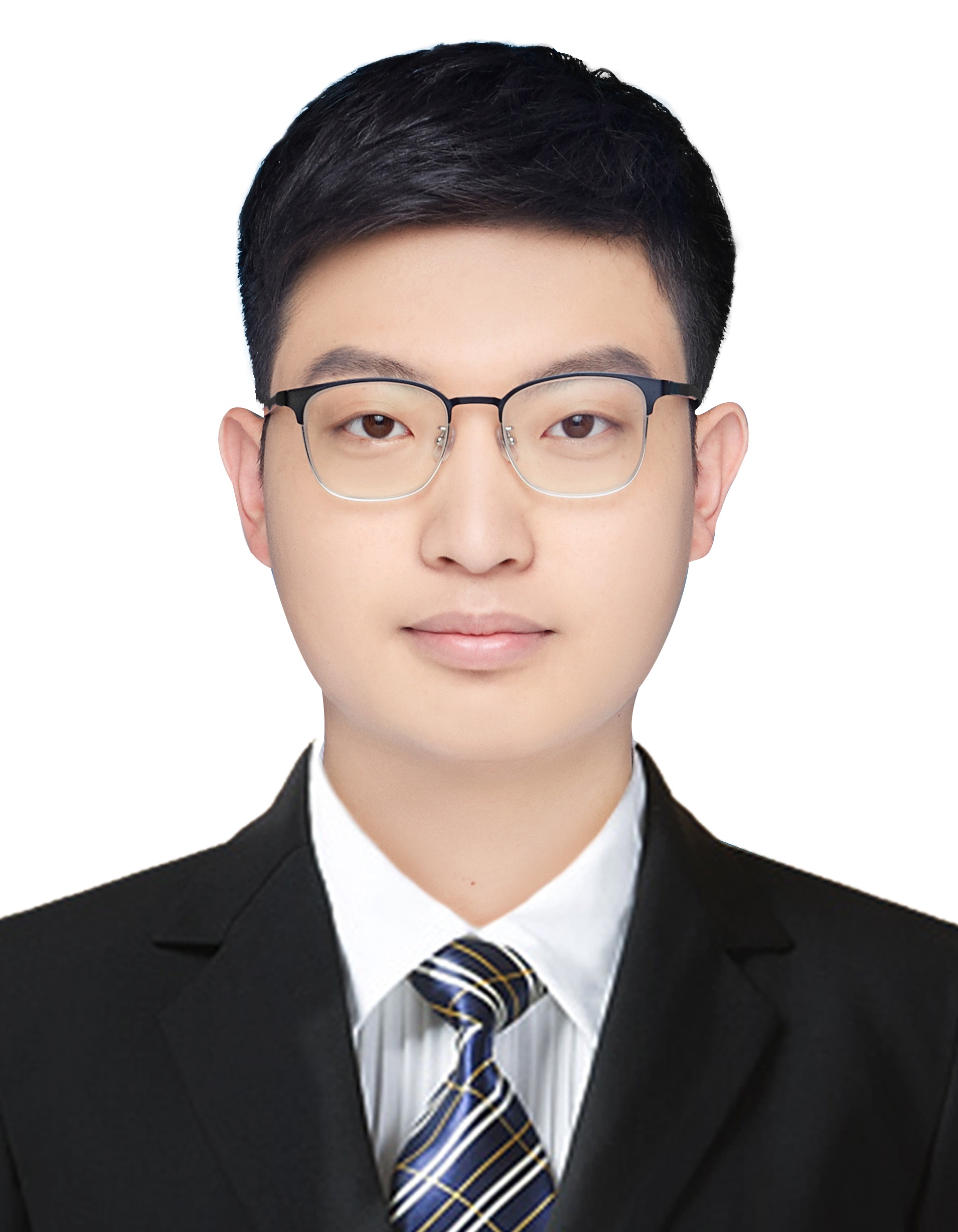}}]{Runyu Yang} received the B.S. and Ph.D. degree degree in electronic information engineering from the University of Science and Technology of China (USTC), Hefei, China, in 2019 and 2025. He is currently a Postdoctoral Fellow of Engineering Science at Simon Fraser University. His research interests include image/video coding and visual feature coding.
	\end{IEEEbiography}
	
	\begin{IEEEbiography}
		[{\includegraphics[width=1in,height=1.25in, clip,keepaspectratio]{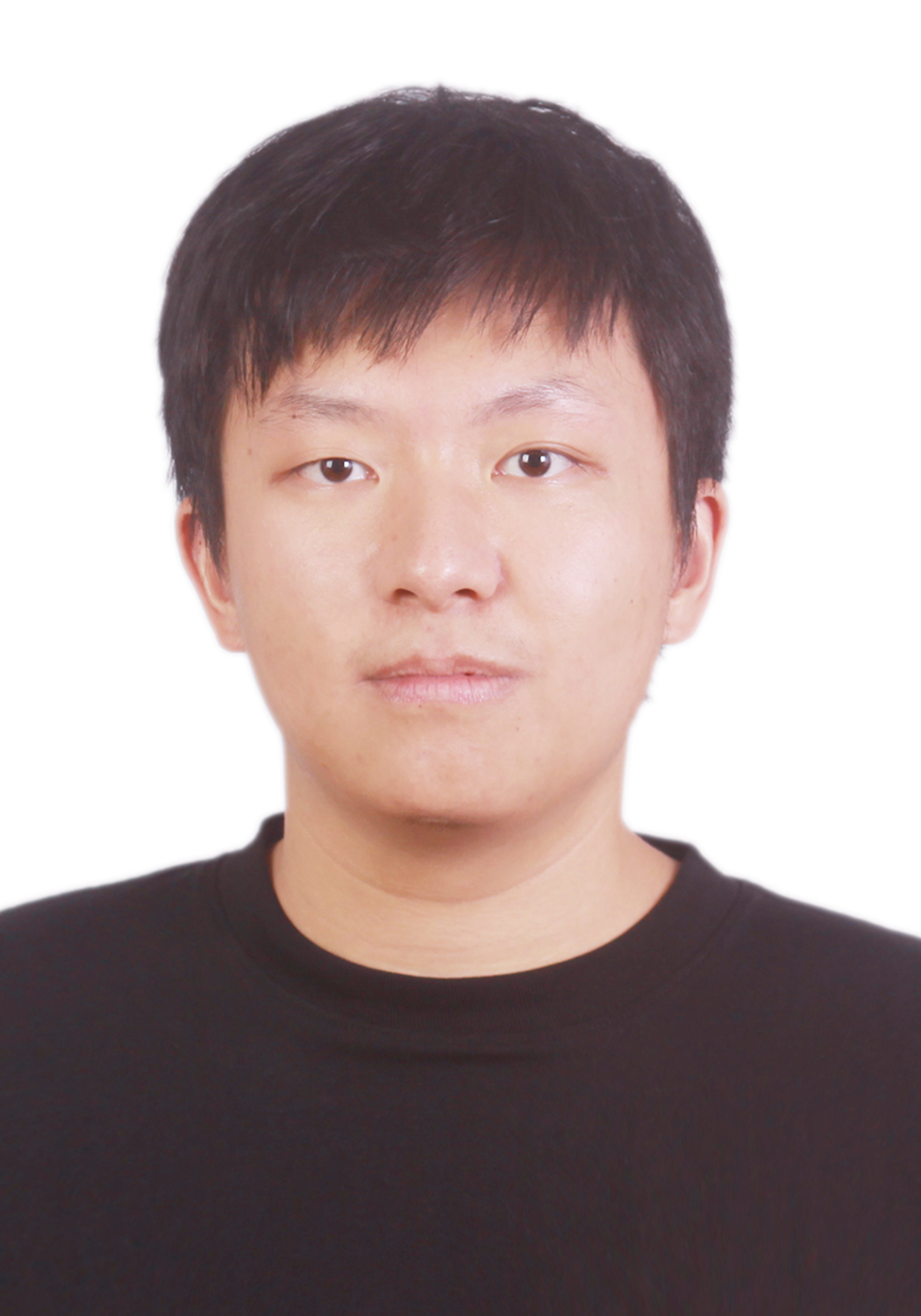}}]{Junqi Liao} received the B.S. degree in electronic information science and technology from Dalian Maritime University (DMU), Dalian, Liaoning, China, in 2021. He is currently working toward the Ph.D. degree with the Department of Electronic Engineering and Information Science, University of Science and Technology of China, Hefei, China. His research interests include image/video coding, reinforcement learning, and computer vision.
	\end{IEEEbiography}
	
	\begin{IEEEbiography}
		[{\includegraphics[width=1in,height=1.25in, clip,keepaspectratio]{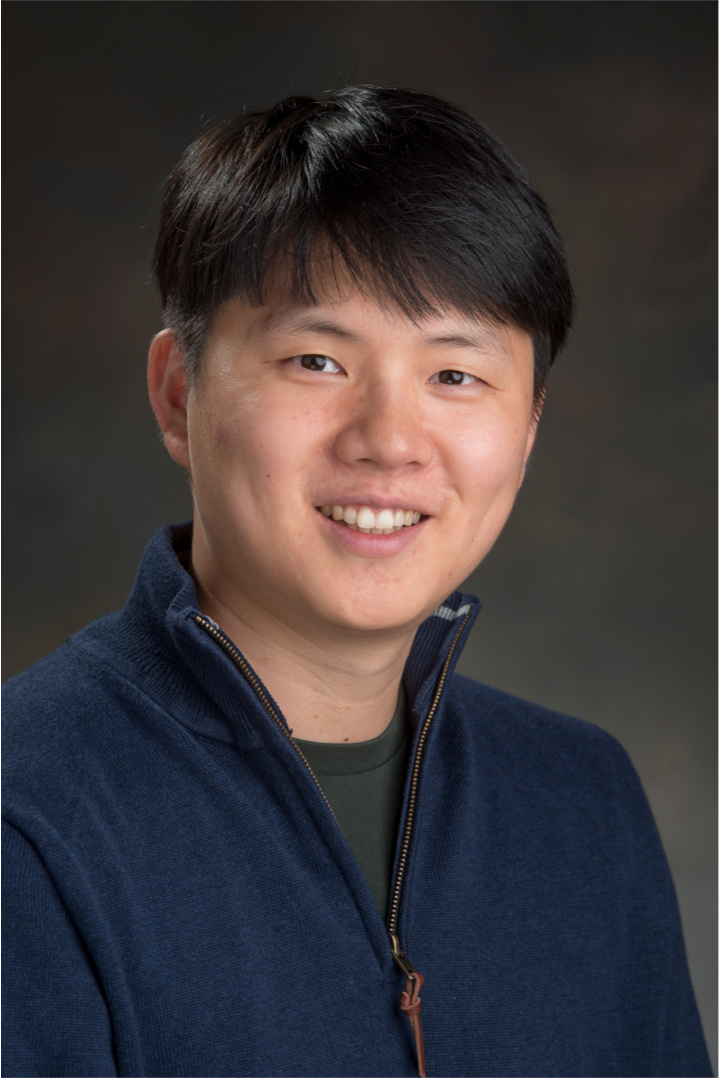}}]{Hyomin Choi} is a Senior Staff Engineer at the AI Lab of InterDigital, Los Altos, CA, USA. He received his Ph.D. degree in Engineering Science from Simon Fraser University, Burnaby, BC, Canada, in 2022. From 2012 to 2016, he was a Research Engineer at the System IC Research Center, LG Electronics, Seoul, Korea. His research interests include end-to-end learning-based image and video coding, video coding for machines, and machine learning for multimedia processing. He currently serves as a Software Coordinator for the MPEG FCM standardization activities. He received the 2017 Vanier Canada Graduate Scholarship, the 2023 Governor General’s Gold Medal from Simon Fraser University, and the 2023 IEEE Transactions on Circuits and Systems for Video Technology (TCSVT) Best Paper Award. He is a member of the IEEE Multimedia Systems and Applications Technical Committee and the IEEE Visual Signal Processing and Communications Technical Committee, and he has served as an Associate Editor for IEEE Transactions on Circuits and Systems for Video Technology (TCSVT).
	\end{IEEEbiography}

	\begin{IEEEbiography}
		[{\includegraphics[width=1in,height=1.25in, clip,keepaspectratio]{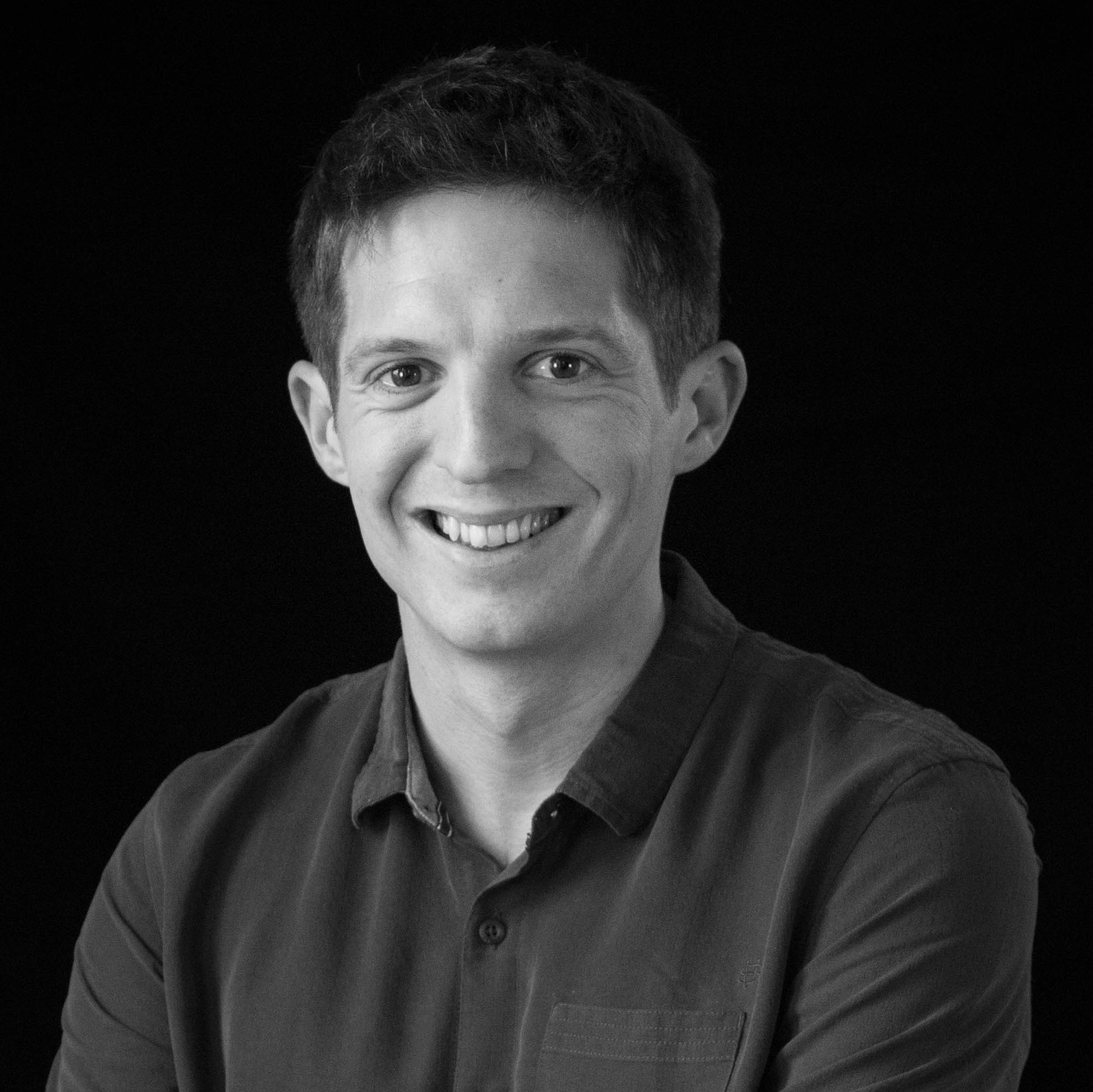}}]{Fabien Racapé} is a Senior Member of Technical Staff with the AI Lab at InterDigital, specializing in video compression, including deep-learning-based compression and coding for machine vision tasks. He serves as an editor for the MPEG Feature Coding for Machines (FCM) standard and has been a contributor to H.266/VVC and MPEG NNC (Neural Network Compression). He received the M.Sc. degree from the Grenoble Institute of Technology in 2008 and the Ph.D. degree from the National Institute of Applied Science (INSA), Rennes, in 2011. He has also served as a TPC member of the Data Compression Conference since 2025. His research interests encompass image/video compression, immersive video coding, perceptual quality, and computer vision.
	\end{IEEEbiography}

	\begin{IEEEbiography}
		[{\includegraphics[width=1in,height=1.25in, clip,keepaspectratio]{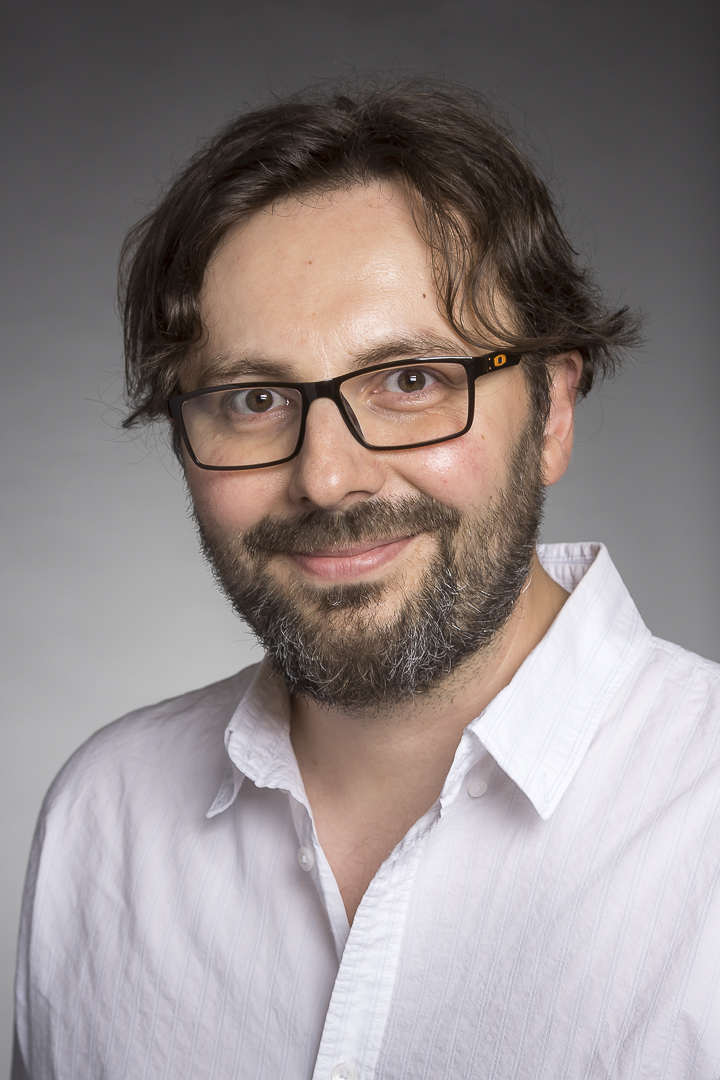}}]{Ivan V. Bajić} is a Professor of Engineering Science at Simon Fraser University, Canada. His research interests include signal processing and machine learning with a focus on multimedia signal processing, compression, and coding for machines. His research has been supported by Canadian, US, and international industry partners and national funding agencies. His work has appeared in leading journals (TPAMI, TIP, TCSVT, TMM) and conferences (CVPR, ICASSP, ICME, ICIP, NeurIPS, ICLR), and has received a number of awards, including the 2023 IEEE TCSVT Best Paper Award, conference paper awards at ICME, ICIP, MMSP, and ISCAS, and other recognitions (best paper finalist, top n\%) at CVPR, ICIP, and Asilomar conferences. He is the past chair of the IEEE Multimedia Signal Processing Technical Committee. He has served on editorial boards of several journals, most recently as Senior Area Editor of IEEE Signal Processing Letters, and had leading roles in several conferences, most recently as TPC Chair of IEEE ICME 2025. 
	\end{IEEEbiography}

	
\end{document}